\documentclass[twocolumn]{aastex6}

\usepackage{amsmath,amssymb,gensymb,times,graphics,morefloats,color}
\usepackage{afterpage, bm}
\usepackage{natbib,hyperref}
\def\kepler{\emph{Kepler}}

\newcommand{\msun}{$\rm M_{\sun}$}
\newcommand{\rsun}{$\rm R_{\sun}$}

\newcommand*\mean[1]{\overline{#1}}

\def\ewha{\mathrm{EW}_{\mathrm{H}\alpha}}
\def\lhalbol{$L_{\mathrm{H}\alpha}/L_{\mathrm{bol}}$}

\def\mk{$M_K$}

\def\mearthmag{M\mkern-2mu Earth}
\def\vsini{$v\sin{i}$}
\def\rossby{$R_o$}

\definecolor{orange}{rgb}{.8,0.4,0}
\definecolor{blue}{rgb}{.0,0.0,1}

\usepackage{soul}

\def\numnew{270}
\def\numall{2202}
\def\numrot{466}

\shorttitle{ }
\shortauthors{Newton et al.}
\begin{document}
\turnoffedit
\title{The H$\alpha$ emission of nearby M dwarfs and its relation to stellar rotation}

\author{Elisabeth R. Newton\altaffilmark{1,2,3}, Jonathan Irwin\altaffilmark{1}, David Charbonneau\altaffilmark{1}, Perry Berlind\altaffilmark{1}, Michael L. Calkins\altaffilmark{1}, Jessica Mink\altaffilmark{1}  \\
}
\altaffiltext{1}{Harvard-Smithsonian Center for Astrophysics, 60 Garden Street, Cambridge, MA 02138, USA}
\altaffiltext{2}{Massachusetts Institute of Technology Kavli Institute for Astrophysics and Space Research, 77 Massachusetts Avenue, Cambridge, MA 02139, USA}
\altaffiltext{3}{NSF Astronomy and Astrophysics Postdoctoral Fellow}

\begin{abstract}

The high-energy emission from low-mass stars is mediated by the magnetic dynamo. Although the mechanisms by which fully convective stars generate large-scale magnetic fields are not well understood, it is clear that, as for solar-type stars, stellar rotation plays a pivotal role. We present \numnew\ new optical spectra of low-mass stars \edit1{in the Solar Neighborhood}. Combining our observations with those from the literature, our sample comprises \numall\ measurements or non-detections of H$\alpha$ emission in nearby M dwarfs. This includes \numrot\ with photometric rotation periods. Stars with masses between 0.1 and 0.6 \msun\ are well-represented in our sample, with fast and slow rotators of all masses. We observe a threshold in the mass--period plane that separates active and inactive M dwarfs. The threshold coincides with the fast-period edge of the slowly rotating population, at approximately the rotation period at which an era of rapid rotational evolution appears to cease. The well-defined active/inactive boundary indicates that H$\alpha$ activity is a useful diagnostic for stellar rotation period, e.g.\ for target selection for exoplanet surveys, and we present a mass-period relation for inactive M dwarfs. We also find a significant, moderate correlation between \lhalbol\ and variability amplitude: more active stars display higher levels of photometric variability. Consistent with previous work, our data show that rapid rotators maintain a saturated value of \lhalbol. Our data also show a clear power-law decay in \lhalbol\ with Rossby number for slow rotators, with an index of $-1.7\pm0.1$. 

\end{abstract}

\section{Introduction}

Solar-type stars show a saturated relationship between rotation and chromospheric or coronal activity: with rotation above a certain threshold, activity maintains a constant value, while at slower spins activity and rotation are correlated. This has been demonstrated using coronal (x-ray) emission \citep[e.g.][]{Pallavicini1981,Vilhu1984, Pizzolato2003, Wright2011}, chromospheric (H$\alpha$ and Ca II) emission \citep[e.g.][]{Wilson1966, Noyes1984, Soderblom1993}, and radio emission from accelerated electrons \citep[e.g.][]{Stewart1988, Slee1989, Berger2006a, McLean2012}. Coronal and chromospheric emission typically result from magnetic heating of the stellar atmosphere, while radio emission is a more direct probe of the magnetic field. The rotation--activity relation is therefore interpreted as resulting from the underlying magnetic dynamo. For solar-type stars, this is generally thought to be the $\alpha\Omega$ dynamo, a product of differential rotation winding up the poloidal magnetic field (the $\Omega$ effect) and subsequent twisting of the now-toroidal magnetic field (the $\alpha$ effect). In the interface $\alpha\Omega$ dynamo, these processes occur at the tachocline, the boundary between the convective and radiative zones. Rotational evolution is also influenced by the magnetic field, due to the coupling of the stellar wind to the magnetic field.

Moving across the M dwarf spectral class, the convective envelope extends deeper into the stellar interior, with theoretical models indicating that stars become fully convective for masses $<0.35$ \msun\ \citep{Chabrier1997}. In stars lacking a tachocline, the interface $\alpha\Omega$ cannot be at play and the mechanisms for generating large-scale magnetic fields in fully-convective stars are not well understood. Nevertheless, Zeeman Doppler imaging reveals that some do indeed have large-scale fields \citep{Donati2008, Morin2010}, and many mid-to-late M-dwarfs have strong signatures of magnetic activity, with emission from the X-ray to the radio \citep[e.g.][]{Berger2010, Stelzer2013}.

Despite the expected difference in magnetic dynamo, the strong connection between rotation and magnetic activity persists in M dwarfs. Consistent with that seen in more massive stars, the rotation-activity relation in M dwarfs is saturated for rapid rotators, and declines with decreasing rotational velocity \citep{Kiraga2007}. This is seen in a wide variety of tracers of magnetic activity, including x-ray flux \citep[]{Stauffer1994, James2000, Wright2011}, Ca H\&K, \citep{Browning2010}, H$\alpha$ emission \citep{Delfosse1998, Mohanty2003, Reiners2012a}, and global magnetic flux \citep{Reiners2009}. The magnetic activity lifetime of low-mass stars is mass-dependent, with spin-down interpreted as the causative factor \citep[e.g.][]{Stauffer1994,Hawley1996,Delfosse1998}. 

Until recently, activity studies for low-mass stars have necessarily had to rely on \vsini\ measurements of rotation rates, which can be obtained only for the most rapidly rotating M dwarfs. The typical \vsini\ survey has a detection threshold of around $3$ km/s, which for a $0.2$ \rsun\ star corresponds to a rotation period of only $3.3$ days. The saturated regime of the rotation--activity relation is seen in stars with detectable rotational broadening, while those without broadening show a range of activity levels. Photometric rotation period measurements, which can probe longer periods, are therefore key to studying the late stages of rotational evolution of low mass stars and the unsaturated rotation--activity relation.  The MEarth Project is a transiting planet survey looking for super Earths around 3000 mid-to-late M dwarfs within 33pc \citep{Berta2012,Irwin2014}. From the MEarth data, we have identified $387$ stars in the northern hemisphere with photometric rotation periods \citep{Newton2016}. Our observations often span 6 months or longer, providing excellent sensitivity to long periods \citep{Irwin2011, Newton2016}. 

\citet{West2015} measured H$\alpha$ emission for $164$ M dwarfs with preliminary rotation periods from MEarth. They found that both the fraction of stars that are active and the strength of magnetic activity declines with increasing rotation period for early M dwarfs. Late M dwarfs were found to remain magnetically active out to longer rotation periods, before both the active fraction and activity level diminished abruptly. In this work, we harness the full MEarth rotation period sample, new optical spectra, and  a compilation of measurements from the literature to undertake an in-depth study of magnetic activity in nearby M dwarfs.

\section{Data}\label{Sec:data}

\subsection{The nearby northern M dwarf sample}

Our sample of M dwarfs is drawn from the MEarth Project, an all-sky survey looking for transiting planets around approximately 3000 nearby, mid-to-late M dwarfs \citep{Berta2012, Irwin2014}. \citet{Nutzman2008} selected the northern MEarth targets from the \citet{Lepine2005a} northern proper motion catalog \edit1{(hereafter the ``nearby northern M dwarf'' sample)}. The sample is composed of all stars with proper motions $>0\farcs15\ \mathrm{yr}^{-1}$, and parallaxes or distance estimates \citep[spectroscopic or photometric;][]{Lepine2005} placing them within 33 pc. The target list for the MEarth transit survey additionally is limited to stars with estimated stellar radii $<0.33$ \rsun, \edit1{but in this work we consider all stars in the nearby northern M dwarf sample}. Note that in the years since our nearby northern M dwarf sample was defined, trigonometric parallaxes have been published for many of these stars, sometimes resulting in revised distances greater than, or estimating radii larger than, \edit1{the limits originally placed on the sample \citep{Dittmann2014}. Thus, stars more massive and more distant than originally intended are included in the nearby northern M dwarf sample.}

MEarth-North is located at the Fred Lawrence Whipple Observatory (FLWO), on Mount Hopkins, Arizona, and has been operational since 2008 September. The observatory comprises eight 40 cm telescopes.
This work utilizes results from MEarth-North and from our further spectroscopic characterization of the sample. Compiled and new rotation period and H$\alpha$ measurements are included in Table \ref{Tab:act_data}, and are described in the following sections. 

\edit1{Our analysis excludes binary stars. Binaries were excluded using the same criteria as in \citet{Newton2016}, which include removing stars with bright, nearby unresolved companions (whether they are background objects or physically associated) and stars that appear over-luminous relative to their colors or spectroscopically-inferred parameters. Binaries are indicated in Table \ref{Tab:act_data}.}

\begin{deluxetable*}{l l l l} 
\tablecaption{\label{Tab:act_data}Magnetic activity measurements and rotation periods for nearby northern M dwarfs (table format)}
\tablecolumns{4}
\tablehead{\colhead{Column} & \colhead{Format} & \colhead{Units} & \colhead{Description}}
\startdata
  1 & A16  & $\cdots$ 		&     2MASS J identifier (numerical part only) \\
  2 & I2  &    h    &      Hour of Right Ascension (J2000) \\
  3 & I2   &   min   &     Minute of Right Ascension (J2000) \\
  4 & F6.3 &   s     &     Second of Right Ascension (J2000) \\
  5 &  A1   &  $\cdots$    &    Sign of the Declination (J2000)\\
  6 & I2  &    deg    &   Degree of Declination (J2000) \\
  7 & I2  &    arcmin &   Arcminute of Declination (J2000) \\
  8 & F6.3 &   arcsec  &  Arcsecond of Declination (J2000)  \\
  9 & F5.3  &  \msun   		& 	Stellar mass\\
  10 & F5.3  &  \rsun    		&  	Stellar radius\\
  11 & F5.3  & $\cdots$ 		&  	$\chi$ value $\times 10^{-5}$ \\ 
  12 & F8.4   & days    			&    	Photometric rotation period\\  
  13 &  A19    & $\cdots$ 		&   ADS bibliography code reference for rotation period \\
  14 & F7.3  &  0.1nm &  	H$\alpha$ EW from this work \\ 
15 & F5.3  &  0.1nm &  	Error in H$\alpha$ EW from this work \\ 
16 & F7.3  &  0.1nm &  	H$\alpha$ EW adopted in restricted sample (with linear correction applied) \\ 
17 & F5.3  &  0.1nm &  	Error in H$\alpha$ EW adopted in restricted sample \\ 
18 & A19    & $\cdots$ &   ADS bibliography code reference for restricted sample H$\alpha$ measurement \\
19 & F6.3  & $\cdots$  &  		\lhalbol\ $\times 10^{-4}$ relative to ``quiescent'' level \\
20 & I1     & $\cdots$ &      Flag indicating upper limit on H$\alpha$ measurement in the unrestricted sample \\ 
21 & F7.3  &  0.1nm &  	H$\alpha$ EW adopted in unrestricted sample \\ 
22 & F5.3 &  0.1nm &  	Error in H$\alpha$ EW in the unrestricted sample \\ 
23 &  A19    & $\cdots$ &   ADS bibliography code reference for unrestricted sample H$\alpha$ measurement \\
24 & I1     & $\cdots$ &    Activity flag flag (0 for active, 1 for inactive) \\
25 & I1    & $\cdots$ & 	Binary flag (0 for no known close companion, 1 for companion)
\enddata
\end{deluxetable*}

\subsection{Incorporating rotation and activity measurements from the literature}\label{Sec:lit}

Our team has undertaken a survey of the literature to gather photometric and spectroscopic data on the nearby northern M dwarfs. We note that our literature survey focused on the mid-to-late M dwarfs that are the targets of the MEarth transit survey, but did not exclude higher mass M dwarfs.

The literature sources for rotation periods are listed in Table \ref{Tab:prot_lit}. 90\% of measurements for $M<0.3$ \msun\ come from \citet{Newton2016}, in which we measured photometric rotation periods for $387$ M dwarfs using photometry from MEarth-North. These measurements supersede those presented previously in \citet{Irwin2014} and \citet{West2015}. We note that the majority of the remaining measurements are from \citet{Hartman2011}. In \citet{Newton2016} we showed excellent agreement between the rotation periods from MEarth and those previously reported in the literature with both measurements. However, \citet{Newton2016} found discrepancies between photometric rotation periods and \vsini\ measurements when \vsini\ values were comparable to the resolution of the spectrum from which they were obtain. Therefore, we do not include \vsini\ measurements in this analysis.

\begin{deluxetable*}{l r r r l}
  \tablecaption{ References for rotation periods
  \label{Tab:prot_lit} }
\tablehead{
\colhead{Reference} &
\colhead{$N_\mathrm{stars}$} }
\startdata
\citet{Krzeminski1969} &      1 \\ 
\citet{Pettersen1980a}  &      1 \\ 
\citet{Busko1980}  &      1 \\
\citet{Pettersen1980} &      1 \\
\citet{Baliunas1983}  &      1 \\ 
\citet{Benedict1998} &      2 \\
\citet{Alekseev1998a} &      1 \\
\citet{Alekseev1998}  &      1 \\
\citet{Robb1999} &      1 \\
\citet{Fekel2000} &      1 \\
\citet{Norton2007} &     10 \\
\citet{Kiraga2007} &      4 \\
 \citet{Engle2009} &      2 \\
\citet{Shkolnik2010} &      5 \\
\citet{Hartman2010} &      6 \\
\citet{Messina2011} &      1 \\
\citet{Hartman2011} &    105 \\
\citet{Kiraga2012} &     11 \\
\citet{Mamajek2013} &      1 \\
\citet{Kiraga2013} &      5 \\
\citet{McQuillan2013} &      6 \\
\citet{Newton2016} &    300
\enddata
\end{deluxetable*}

The sources for H$\alpha$ measurements are listed in Table \ref{Tab:ewha_lit}. The table and the discussion below includes the new measurements we make in this work; our observations are discussed in \S\ref{Sec:observations} and our EW measurements in \S\ref{Sec:ew}. H$\alpha$ measurements are derived from spectra obtained using instruments with various spectral resolutions, and some analyses used different definitions of, or different means of calculating, H$\alpha$ EW. Additionally, not all literature sources report values when a star was considered to be inactive. We note that for H$\alpha$ EWs from \citet{Alonso-Floriano2015} we treat H$\alpha$ EWs reported as exactly $0.0\pm0.2$ as upper limits. When multiple measurements of H$\alpha$ are available, we adopt one measurement rather than averaging those available. 

In this paper, we use H$\alpha$ measurements in two different ways, and for each purpose we use a different criterion to choose which of the available H$\alpha$ measurements to adopt:

If we use the value of the EW measurement, we only adopt H$\alpha$ measurements from literature sources for which we are able to ensure good agreement with our measurements through comparison of overlapping samples across all activity levels. We also apply a linear correction to the EWs from each literature survey, calibrated using the overlapping stars, and thus require that a linear correction is sufficient to account for differences (high order polynomials may misbehave outside of the calibrated range). We do not include upper limits. We were able to verify agreement with \citet{Gizis2002}, \citet{Gaidos2014}, and \citet{Alonso-Floriano2015}. When measurements are available from more than one of these sources or our own work, we select the measurement obtained from the highest resolution spectrum.We refer to the sample of H$\alpha$ measurements selected (and corrected) in this way as the ``restricted sample.''

If we consider only whether a star is active or inactive, we do not restrict the literature sources from which we adopt H$\alpha$ measurements, and we include upper limits. We adopt the measurement from the restricted sample if possible. If a measurement is not available in the restricted sample, we adopt a measurement from any other available source with preference given to the measurement obtained at the highest spectral resolution. We then use the EW value to determine whether a star is active or not. If the measurement is an upper limit, we consider the star to be inactive. Otherwise, we adopt $-1$ \AA\ as our active/inactive boundary. Though \citet{West2015} found that a threshold of $-0.75$ \AA\ was appropriate for spectra obtained using the same instrument and settings as we use in this work, the threshold most commonly used in the literature sources we gathered was $-1$ \AA. There are not many stars with EWs between $-0.5$ and $-1.5$ \AA, so choosing a different boundary does not result in significant differences. We refer to the sample of H$\alpha$ measurements selected in this way as the ``unrestricted sample.''

\begin{deluxetable*}{l r r r l}
  \tablecaption{References for H$\alpha$ equivalent widths
 \label{Tab:ewha_lit}}
\tablehead{
\colhead{Reference} &
\colhead{Resolution} &
\colhead{$N_\mathrm{restr}$\tablenotemark{a}} &
\colhead{$N_\mathrm{unrestr}$\tablenotemark{b}} &
\colhead{Restr. sample correction\tablenotemark{c}}}
\startdata
\citet{Reid1995}\tablenotemark{d}            &   2000 &    0 &  343 & N/A \\ 
\citet{Martin1996}            &   2340 &    0 &    1 & N/A \\
\citet{Hawley1996}\tablenotemark{d}            &   2000 &    0 &   31 & N/A \\
\citet{Stauffer1997a}           &  44000 &    0 &    1 & N/A \\ 
\citet{Gizis1997b}           &   2000 &    0 &    7 & N/A \\ 
\citet{Tinney1998}            &  19000 &    0 &    1 & N/A \\ 
\citet{Gizis2000}            &   1000 &    0 &    8 & N/A \\
\citet{Cruz2002}            &   1400 &    0 &   14 & N/A \\
\citet{Gizis2002}           &  19000 &  428 &  428 & $a=1.01$, $b=-0.162$ \\
\citet{Reid2002}            &  33000 &    0 &   19 & N/A \\
\citet{Lepine2003}           & multiple &    0 &    1 & N/A \\
\citet{Mohanty2003}            &  31000 &    0 &    3 & N/A \\
\citet{Bochanski2005}            &  multiple &    0 &    4 & N/A \\
\citet{Phan-Bao2006}           &   1400 &    0 &    5 & N/A \\ 
\citet{Riaz2006}           &   1750 &    0 &   19 & N/A \\ 
\citet{Reid2007}           &   1300 &    0 &   29 & N/A \\ 
\citet{Reiners2007}           &  31000 &    0 &    1 & N/A \\
\citet{Reiners2008}            &  31000 &    0 &    3 & N/A \\
\citet{Lepine2009}           &    multiple &    0 &    2 & N/A \\
\citet{Shkolnik2009}            &  60000 &    0 &   29 & N/A \\
\citet{Reiners2010}            &  31000 &    0 &   10 & N/A \\
\citet{West2011}           &   1800 &    0 &   12 & N/A \\
\citet{Lepine2013}           &   2000, 4000 &    0 &   13\tablenotemark{e} & N/A \\ 
\citet{Gaidos2014}            &    $\sim$1200 &  582 &  582 & $a=1.00$, $b=-0.253$ \\
\citet{Ivanov2015}            &    $\sim$1000 &    0 &    1 & N/A \\ 
\citet{Alonso-Floriano2015}            &   1500 &   99 &  179\tablenotemark{f} & $a=1.01$, $b=0.866$ \\
\citet{West2015} &       3000 &  0 &    0 & N/A\tablenotemark{g} \\ 
This Work                       &    3000 &  456 &  456 & N/A
\enddata
\tablenotetext{a}{$N_\mathrm{restr}$ is the number of stars included in the restricted sample, for which a linear correction to the H$\alpha$ EW was applied to ensure agreemenet between literature values and those measured in this work. This sample is used when we consider the value of the H$\alpha$ EW.}
\tablenotetext{b}{$N_\mathrm{unre}$ is the number of stars included in the unrestricted sample. This sample includes entries where only a limit of H$\alpha$ EW was reported, or for which agreement with our measurements could not be assured. The ``active/inactive'' flag is based on the unrestricted sample.}
\tablenotetext{c}{The coefficients for the linear correction applied to produce values in the restricted sample: $\mathrm{EW}_\mathrm{restr} = a\times\mathrm{EW}_\mathrm{lit}+b$.}
\tablenotetext{d}{\citet{Reid1995} and \citet{Hawley1996} report H$\alpha$ indices; we use EWs from I.N. Reid's website \url{http://www.stsci.edu/~inr/pmsu.html}.}
\tablenotetext{e}{We have opted to have values from \citet{Gaidos2014} supersede those from \citet{Lepine2013}.}
\tablenotetext{f}{\citet{Alonso-Floriano2015} includes upper limits, thus the number of stars from this source that are in the unrestricted sample exceeds the number that are in the restricted sample.}
\tablenotetext{g}{We have re-reduced and re-analyzed the data first presented in \citet{West2015} to ensure consistent results.}
\end{deluxetable*}

\subsection{New optical spectra from FAST}\label{Sec:observations}

We obtained new optical spectra for \numnew\ M dwarfs.
We used the FAST spectrograph on the $1.5$ m Tillinghast Reflector at FLWO. We used the $600$ lines mm$^{-1}$ grating with a $2\arcsec$ slit, resulting in approximately $2$ \AA\ resolution ($R=3000$) over $2000$ \AA. We used a tilt setting of 752, corresponding to a central wavelength of $6550$ \AA, to obtain spectra covering $5550-7550$ \AA.

The data were reduced using standard IRAF long-slit reductions. Using calibration exposures taken at each grating change, the 2D spectra were rectified, bias-subtracted and flat-fielded. The wavelength calibration was determined from a HeNeAr exposure taken immediately after each science observation. A boxcar was used to extract 1D spectra, with linear interpolation to subtract the sky. We did not clean cosmic rays or weight pixels in the cross-dispersion direction, because we found that these processes could suppress the resulting H$\alpha$ EW by a few percent for strong emission lines. We used spectrophotometric standards to perform a relative flux calibration. 

In \citet{West2015}, we presented spectra of $238$ additional M dwarfs including measurement of H$\alpha$ EWs. These spectra were obtained using the same instrument and settings, but extraction included cleaning and weighting. The difference is a decrease in the EWs of about $3\%$ for some strongest emission lines. To ensure consistent analysis, we re-reduce these spectra using the steps outlined above. \edit1{The stars with observations from \citet{West2015} and the new observations we present here do not overlap, which brings the number of M dwarfs with FAST spectra obtained by members of our team to $508$.}

In \S\ref{Sec:ew} we discuss our analysis of these data to measure chromospheric activity. We refer to Appendix \ref{Sec:rv} for a thorough discussion of measuring accurate absolute kinematic radial velocities from these spectra.

\subsection{Newly identified multiple systems}

We have identified several new M dwarf binary systems. Prior to this work, 2MASS J12521285+2908568 (LP 321-163) was identified as a spectroscopic double-lined system by our team using TRES on the $1.5$ m Tillinghast Reflector at FWLO.

In the course of our analysis of the FAST spectra, we identified three M dwarfs with unusually broad H$\alpha$ emission lines that had no previous identification as multiple systems. We obtained high resolution optical spectra using TRES, which confirmed our suspicions. We found that 2MASS J09441580+4725546 and 2MASS J16164221+5839432  (G 225-54) are spectroscopic triple-lined systems, while 2MASS J11250052+4319393 (LHS 2403) is double-lined.

\edit1{A TRES spectrum of 2MASS J20245996+0225569 (LP 635-10) revealed that it is also a spectroscopic double-lined system. We pursued high resolution follow-up spectroscopy of this object when we noticed that it showed moderate magnetic activity (EW$=-1.7$\AA), unusual for a slowly rotating M dwarf (see \S\ref{Sec:inactive}). The TRES spectrum does not show H$\alpha$ emission.}

\subsection{Revisions to literature rotation periods}

We note revisions to several rotation periods from the literature, which we investigated further due to unusual activity levels (see \S\ref{Sec:inactive}). For 2MASS J08111529+3607285 (G 111-60), \citet{Hartman2011} report a period of $3.29$ days. We have determined that this period is for a nearby RS CVn variable which was blended in their photometry.

\edit1{For 2MASS J08175130+3107455 (G 90-52), \citet{Hartman2011} report a period of $0.97$ days, and flag it as a possible blend. A high resolution spectrum obtained with TRES did not show rotational broadening, suggesting either that the sub-one day photometric modulation is an alias of a longer period, or that it derives from the nearby companion.}
 
\edit1{Accordingly, the rotation periods for these stars not been included.}

\section{Chromospheric activity}\label{Sec:ew}

\edit1{Prior to calculating EWs, we perform relative flux calibration and shift the spectra to zero radial velocity. We measure radial velocities by matching each spectrum to a zero-velocity standard from SDSS \citet{Bochanski2007a}. We adjust for the known $7.3$ km/s offset in the SDSS absolute RV rest frame \citet{Yanny2009}, which we note was not corrected for in \citet{West2015}. We estimate a precision of $8$ km/s and identify the potential for $5$ km/s systematic offsets between spectral types. Further details on our RV method and tests for accuracy and precision can be found in \S\ref{Sec:rv}.}

\subsection{H$\alpha$ EWs}

We measure H$\alpha$ EWs using the definition of \citet{West2011}. The continuum $F_c$ is given by the mean flux across two regions to either side of the H$\alpha$ feature, $6500-6550$ \AA\ and $6575-6625$ \AA. The EW is:
\begin{equation}
\mathrm{EW} = \sum{ 1-\frac{F(\lambda)}{F_c} \delta\lambda}
\end{equation}
The limits of the integral are $6558.8-6566.8$ \AA. We sum the flux within the feature window, using fractional pixels as necessary and assuming pixels are uniformly illuminated. EWs for H$\alpha$ in emission are given as negative values.

\begin{figure}
\includegraphics[width=0.9\linewidth]{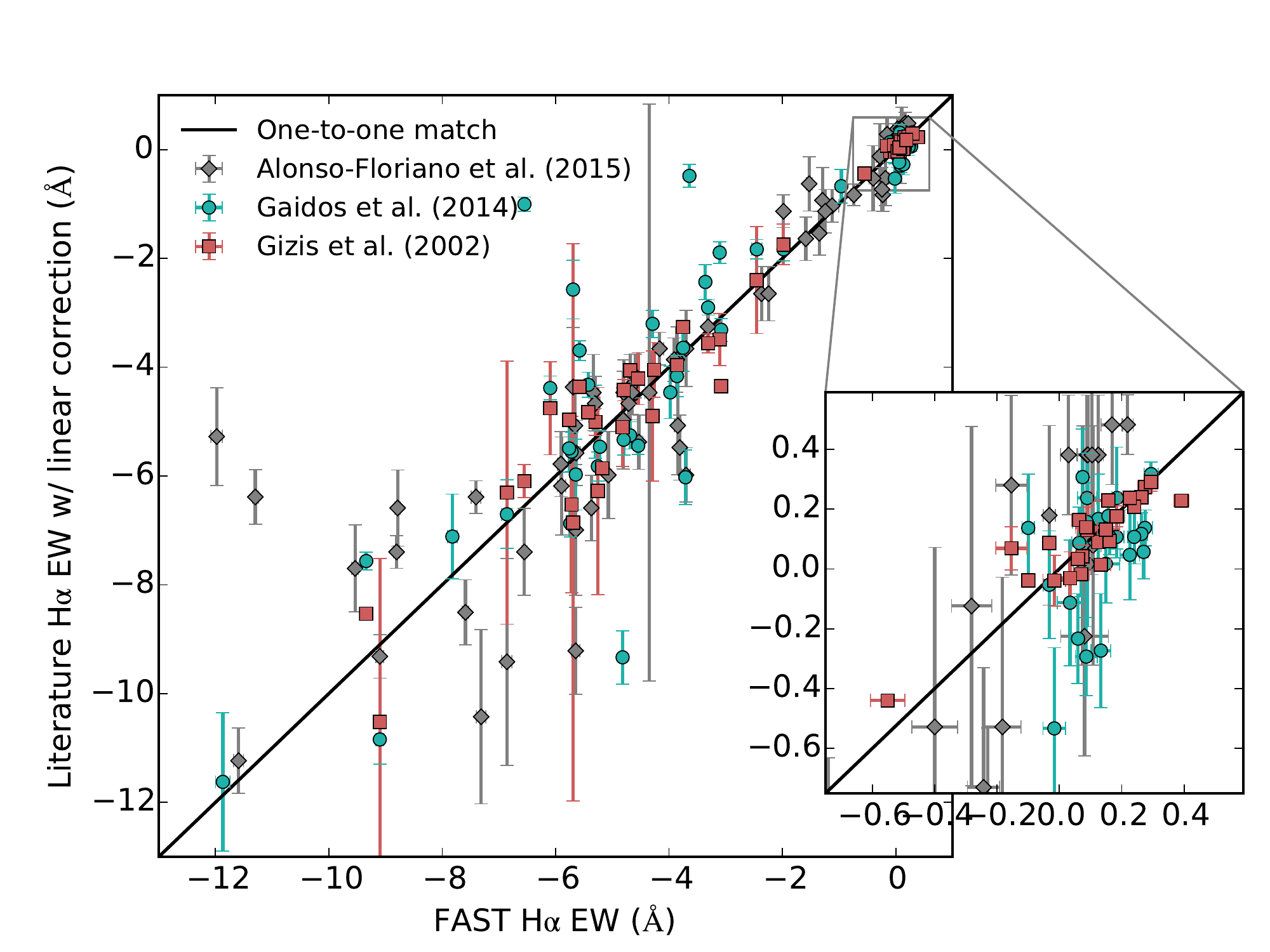}

\caption{Comparison of H$\alpha$ EWs with measurements from this work and from the literature, for the three surveys included in our restricted sample. The comparison to \citet{Alonso-Floriano2015} is shown as gray diamonds, to \citet{Gaidos2014} as teal circles, and to \citet{Gizis2002} as orange squares. The inset highlights the inactive stars. We have applied a linear correction to literature values by fitting the overlapping sample.}\label{Fig:ewcomp}
\end{figure}

Figure \ref{Fig:ewcomp} compares our EW measurements to those from \citet{Gizis2002}, \citet{Gaidos2014}, and \citet{Alonso-Floriano2015}, the three surveys which are included in our restricted sample of H$\alpha$ EWs (see \S\ref{Sec:lit}). There are $52$ stars in our sample with measurements from \citet{Gizis2002}, and $70$ from \citet{Alonso-Floriano2015}, with no stars in either sample which have discrepant identification as active or inactive.
There are $60$ stars with measurements both from our sample and from \citet{Gaidos2014}, including $33$ with H$\alpha$ detected in emission in our data. Using a limit of $-1$ \AA, there are two stars identified as inactive in \citet{Gaidos2014} that have strong H$\alpha$ in our spectra. For 2MASS J17195298+2630026, which is a member of a close visual binary, our detection of strong H$\alpha$ emission agrees with previous measurements from \citet[][$-8.4$ \AA]{Reid1995}, \citet[][$-8.2$ \AA]{Alonso-Floriano2015}, and \citet[][$-5.9$ \AA]{Gizis2002}. For 2MASS J03524169+1701056, the only other value is from \citet{Reid1995}, who get $-1.23$ \AA; this is between our value ($-3.2$ \AA) and the non-detection from \citet{Gaidos2014}. \edit1{This case could result from intrinsic stellar variability, though the amount of variability required is larger than typical.}

Considering only the active stars and removing measurements that deviate by $>3$ \AA, the standard deviation in EW measurements is $0.7$ \AA\ for \citet{Gizis2002}, and $1$ \AA\ for both \citet{Gaidos2014} and \citet{Alonso-Floriano2015}. This is similar to the intrinsic variability of around $0.8$ \AA\ seen in time-resolved measurements \citep[][]{Lee2010, Bell2012}.

\subsection{H$\alpha$ relative to quiescence}

\begin{figure}
\includegraphics[width=0.9\linewidth]{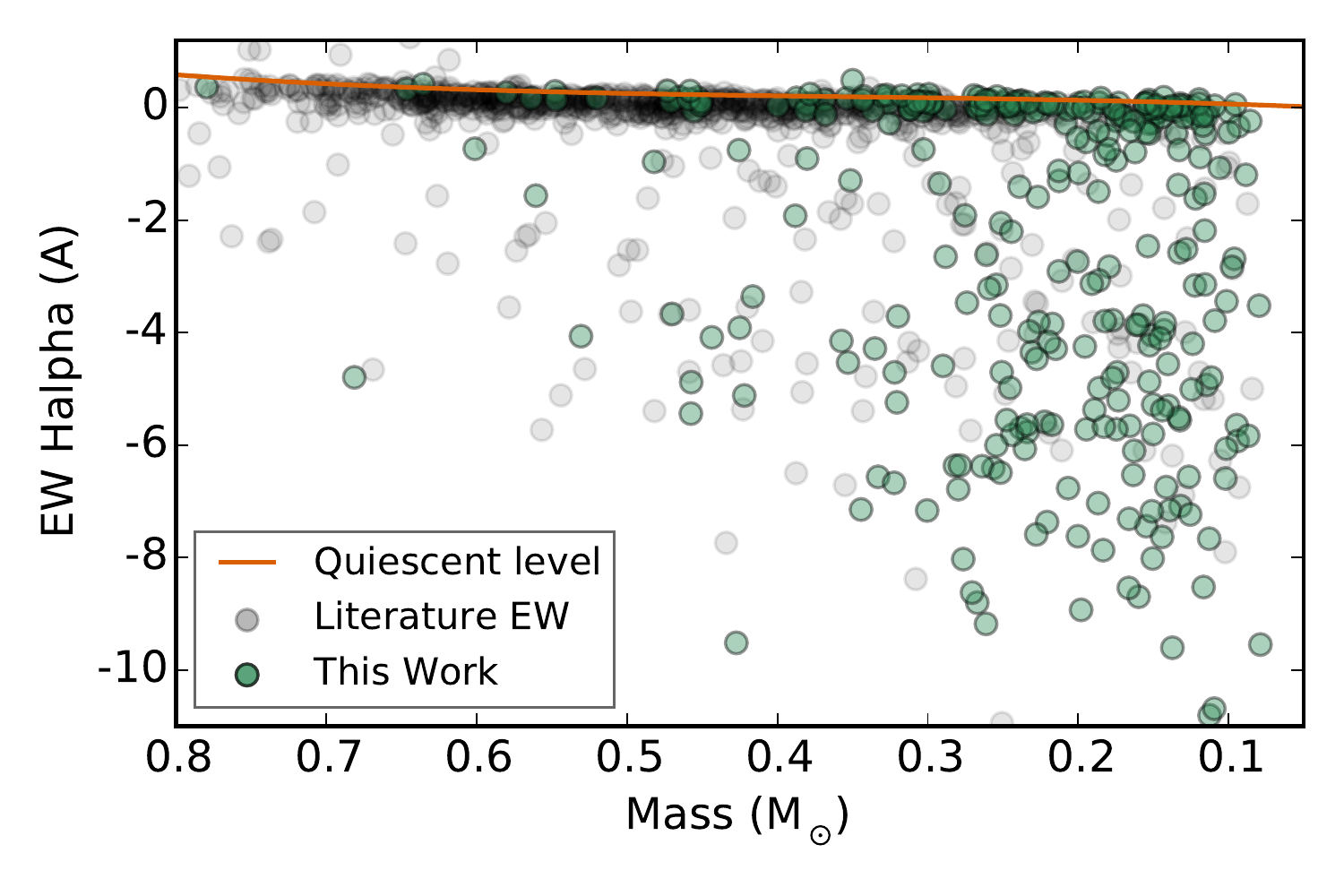}

\caption{H$\alpha$ EW as a function of stellar mass. Masses are estimated from the absolute $K$ magnitudes as in \citet{Newton2016} and have a typical error of $10\%$. Green circles show measurements obtained in this work; gray circles are drawn from the literature. Only stars in our restricted sample (\S\ref{Sec:lit}) are shown. The orange line indicates the ``quiescent'' activity level relative to which we measure H$\alpha$ EWs when calculating \lhalbol.}\label{Fig:quiescent}
\end{figure}

\citet{Wilson1957} found that the strength of the Ca K line reversal could be used as a luminosity indicator in late-type stars.  \citet{Kraft1964} identified the Wilson-Bappu effect using H$\alpha$ absorption in late-type stars, and \citet{Stauffer1986} demonstrated the effect in M dwarfs. The trend is one of smaller absorption EWs for redder stars. Figure \ref{Fig:quiescent} shows a clear mass-dependent envelope to H$\alpha$ absorption strengths, consistent with these findings: the maximum amount of absorption decreases for smaller mass M dwarfs. We correct our EWs so that we measure the amount of emission above this maximum absorption level. 

We fit the H$\alpha$ envelope as a function of mass by iteratively rejecting stars with EWs more negative (higher emission) than the best fit. We consider this to be the ``quiescent'' level, and when calculating \lhalbol\ we correct our measured EWs based on their estimated stellar masses. 
\begin{align}
\mathrm{EW}_\mathrm{quiescent}/\mathrm{\AA} =& -0.03918 \\ \nonumber 
&+ 1.319\times (M_*/M_\odot)  \nonumber \\
&- 2.779\times (M_*/M_\odot)^2 \nonumber \\
&+2.635\times (M_*/M_\odot)^3
\end{align}
\begin{align}
\mathrm{EW}_\mathrm{relative}= \mathrm{EW}_\mathrm{measured} - \mathrm{EW}_\mathrm{quiescent}
\end{align}

Weak activity levels are thought to first strengthen absorption as the $n=2$ state\footnote{H$\alpha$ of course corresponding to the transition between the $n=2$ and $n=3$ energy levels.} is populated, with increasing activity eventually strengthening emission as the $n=3$ state becomes populated \citep{Cram1979, Stauffer1986}. Our data are not sufficient to distinguish whether a weakly active star is in the earlier stage (strengthening absorption) or the latter stage (strengthening emission) of chromospheric heating. By measuring EWs relative to the maximum amount of absorption, we are assuming that all M dwarfs in our sample are active enough to have at least reached this maximum absorption level.

\subsection{$\chi$ and relative \lhalbol}\label{Sec:chi}

The H$\alpha$ luminosity, \lhalbol, is commonly used to enable comparison between stars of different intrinsic luminosities. Calculation of intrinsic H$\alpha$ luminosity requires absolutely flux-calibrated spectra. Accurate photometry in the wavelength region covered by our FAST spectra is not widely available for our sample, and absolute flux calibration is beyond the scope of the present work. The $\chi$ factor is commonly used in this circumstance \citep{Walkowicz2004, West2008a}. The $\chi$ factor is derived from photometric colors, and \lhalbol\ is then easily calculated: \lhalbol$=\ewha\times\chi$. We adopt $\chi$ factors from \citet{Douglas2014}, who found significant differences compared to previous work. We refer the reader to \citet{Douglas2014} for a thorough discussion.

The \citet{Douglas2014} $\chi$ factor is presented as a function of $r'-J$ or $i-J$. Neither $r'$ nor $i$ is widely available for our sample, which even if they are within the SDSS footprint are typically saturated.  In \citet{Dittmann2016}, we calibrated the MEarth photometric system, and presented $\mearthmag$ magnitudes for $1507$ M dwarfs. \citet{Dittmann2016a} obtained absolute $griz$ Sloan photometry for 150 MEarth M dwarfs using the filters on the FLWO $1.2$ m ($48$ in.). We use these data to derive the conversion between $\mearthmag-J$ and $i_{48}-J$: 
\begin{equation}\label{Eq:ijcol}
i_{48}-J = 1.391 \times (\mearthmag - J) + 0.139
\end{equation}
The MAD of this conversion is $0.03$ mag and the standard deviation is $0.05$, the latter of which we adopt as the error on $i_{48}-J$ colors calculated in this way. The difference between $i_{48}$ and $i_\mathrm{SDSS}$ is likely small\footnote{We note that this is not the case for $r$ magnitudes, where the filter edge may overlap with a  sharp spectroscopic feature, as discussed in \citet{Dittmann2016a}}, and we do not make additional corrections. \edit1{We discard estimated $i_{48}-J$ colors that are $>2\sigma$ outliers in the mass--($i_{48}-J$) plane.}

Not all stars in our sample have a $\mearthmag$ magnitude, but $I_C$ magnitudes are sometimes available. We fit a relation between $I_C-J$ and $i_{48}-J$, using the $i_{48}-J$ colors previously inferred from $\mearthmag-J$ as the basis for the fit:
\begin{align}
i_{48}-J =&+ 0.142 \\
&+ 1.522\times (I_C-J) \\ \nonumber
& -0.123\times (I_C-J)^2 \\ \nonumber
\end{align}
The MAD of this conversion is $0.016$ mag and the standard deviation is $0.025$ mag, the latter of which we adopt as the error on $i_{48}-J$ calculated in this way.

We then calculate $\chi$ using the relation presented in the Appendix of \citet{Douglas2014}. Figure \ref{Fig:chi-sdss} demonstrates excellent agreement between our estimated $\chi$ values and the mean values calculated by \citet{Douglas2014} as a function of spectral type. Spectral types for nearby M dwarfs are taken from the literature. Intrinsic scatter in $\chi$ as a function of spectral type is apparent, reflective of the imperfect mapping between spectral type and color for M dwarfs.

\begin{figure}
\includegraphics[width=0.9\linewidth]{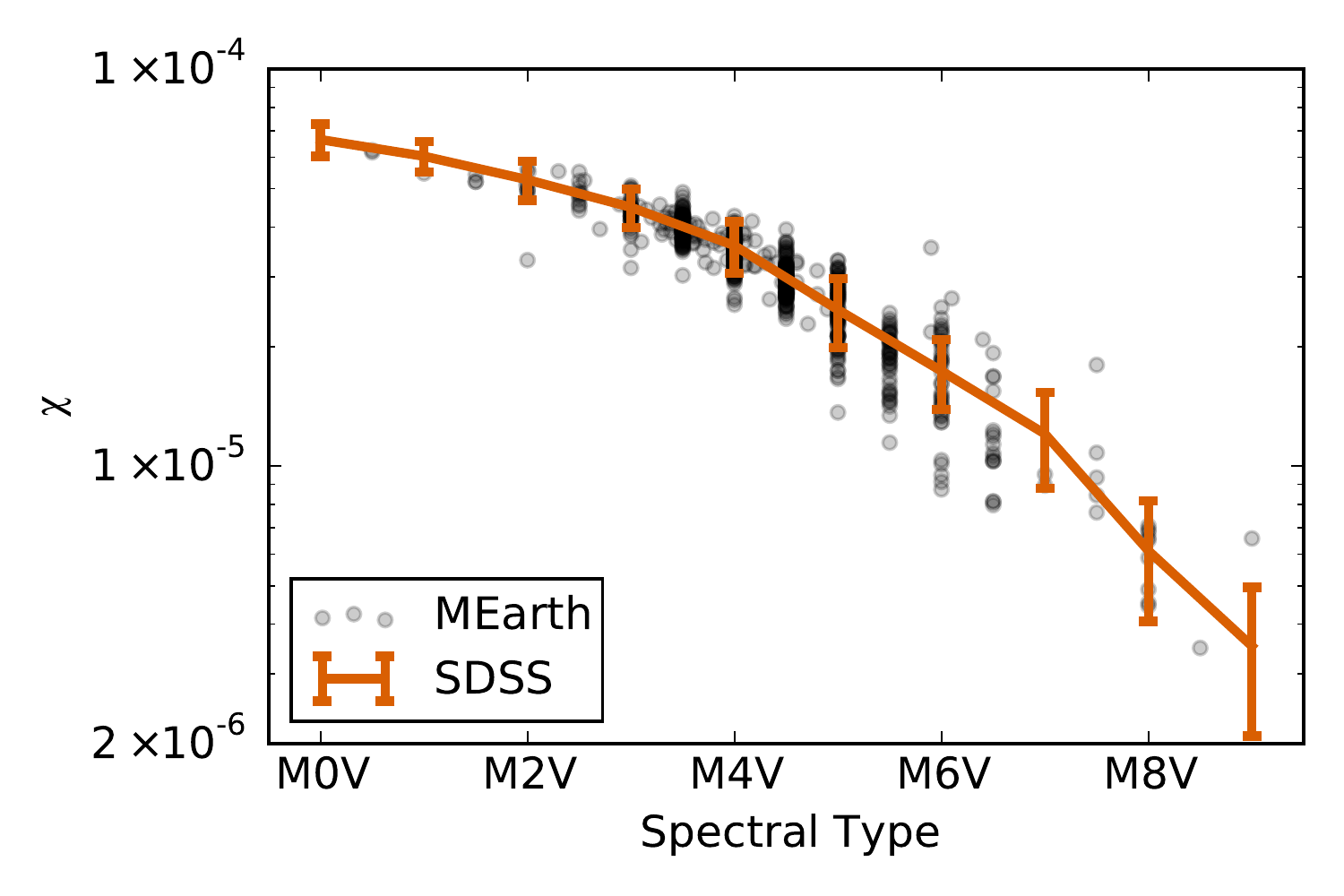}

\caption{$\chi$ values, which are used to infer \lhalbol\ from H$\alpha$ EW, plotted versus spectral type. In black are the values we infer for stars in our sample using the new calibration from \citet{Douglas2014}, with $i-J$ colors estimated as described in \S\ref{Sec:chi}. Optical spectral types are drawn from the literature. In orange are the mean values \citet{Douglas2014} measure for M dwarfs in SDSS, with error bars indicating the standard deviation in each bin. \label{Fig:chi-sdss}}
\end{figure}

\begin{figure}
\includegraphics[width=0.9\linewidth]{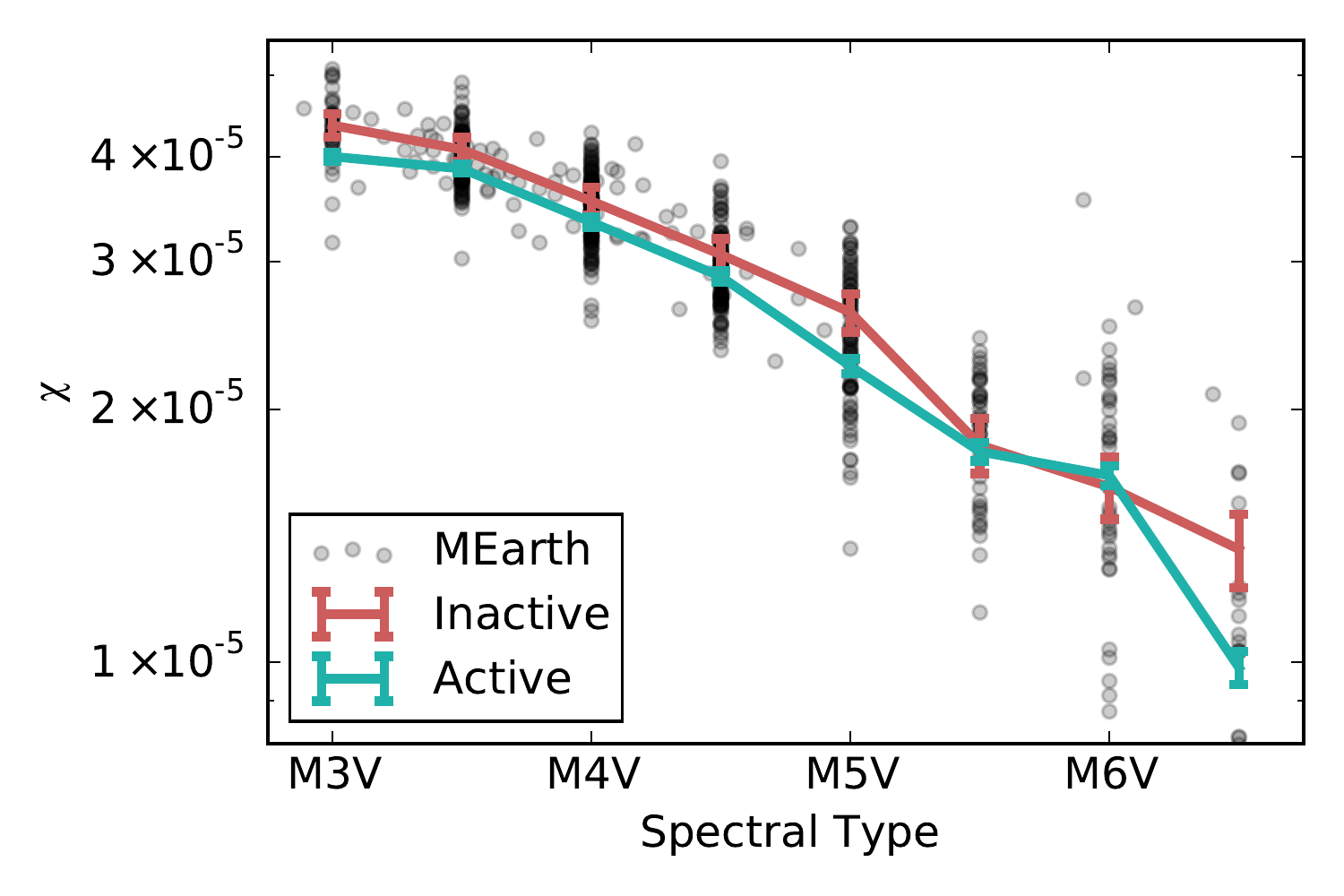}

\caption{$\chi$ values versus spectral type. In black are the values we infer for stars in our sample using the new calibration from \citet{Douglas2014}, with $i-J$ colors estimated as described in \S\ref{Sec:chi}. Optical spectral types are drawn from the literature. The mean values of $\chi$ for active stars in our sample are in cyan and the inactive stars in red, with the error bars indicating the standard error on the mean.  \label{Fig:chi-active}}
\end{figure}

One potential concern is whether $\chi$ depends on the level of H$\alpha$ emission itself. We see a small but statistically significant difference between active and inactive M dwarfs, in that active M dwarfs have slightly lower mean $\chi$ values (Figure \ref{Fig:chi-active}). For M3V--M5V, where we have sufficient numbers of both samples for a meaningful comparison, the difference and standard error, expressed as a percentage of the mean value for inactive stars, is about $5.5\pm1.5\%$. Redder stars have smaller $\chi$ values, so this is equivalent to the more active stars having redder colors at a given spectral type. Such an effect has been seen in previous works, e.g. \citet{Hawley1996}. If we assume that, at a given spectral type, stars with larger H$\alpha$ EW are redder, \lhalbol\ will scale less than linearly with H$\alpha$ EW. Whether this is a relevant astrophysical effect or a systematic one requires further investigation, but in either case the intrinsic scatter dominates.

We only calculate \lhalbol\ for the stars in the restricted sample (defined in \S\ref{Sec:lit}).

\section{Results}

We look at the relationship between activity and rotation as a function of stellar mass. Our photometric rotation periods allow us to probe longer rotation periods than typically accessible for low-mass stars. We use the empirically calibrated relationship between mass and absolute $K$ magnitude (calculated using  trigonometric parallaxes only) to infer stellar mass \citep{Delfosse2000}, which we modify as discussed in \citet{Newton2016} to allow extrapolation. We have excluded known binaries from this analysis, as discussed \S\ref{Sec:data}.

\subsection{The active/inactive boundary}\label{Sec:inactive}

\begin{figure}
\includegraphics[width=\linewidth]{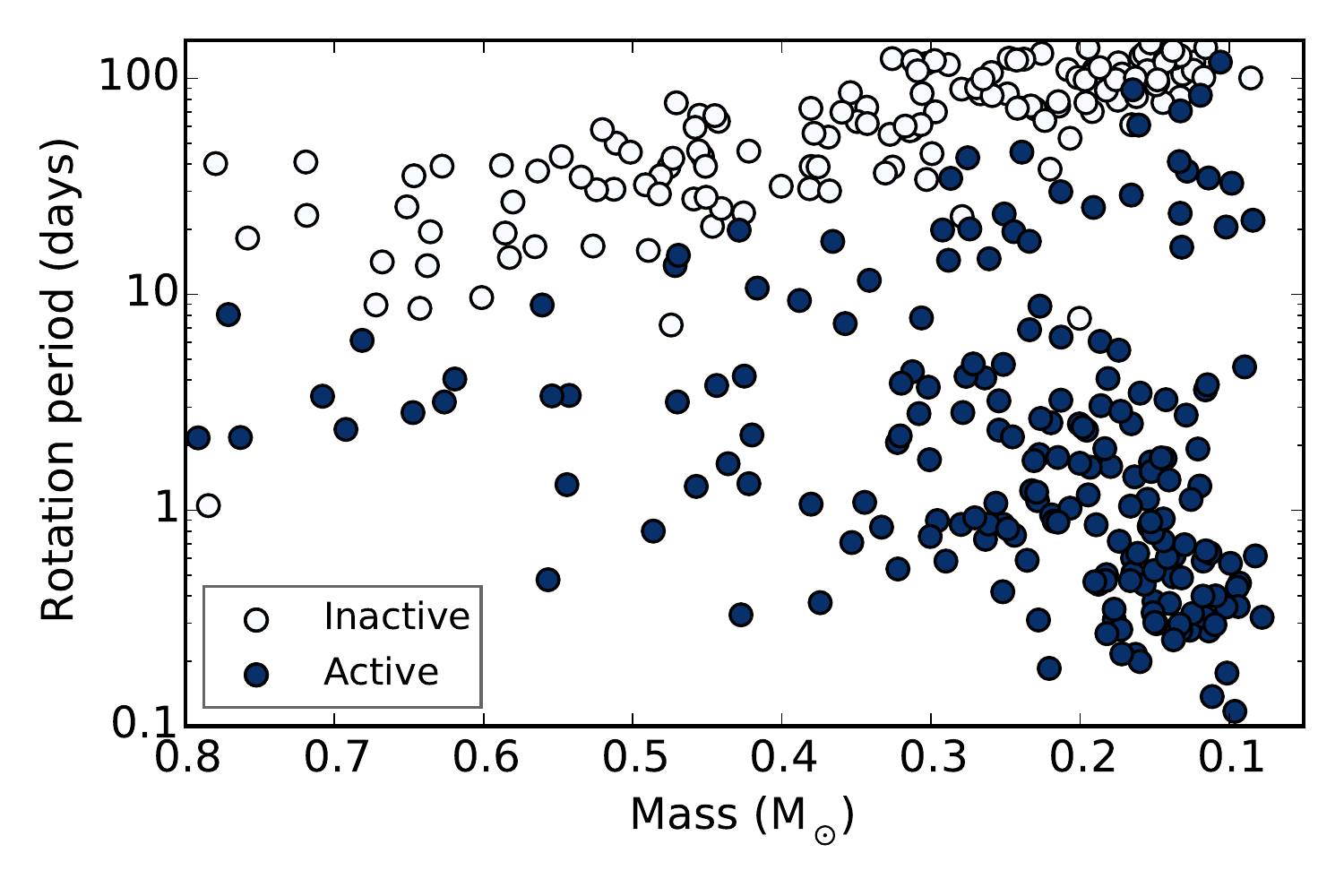}

\includegraphics[width=\linewidth]{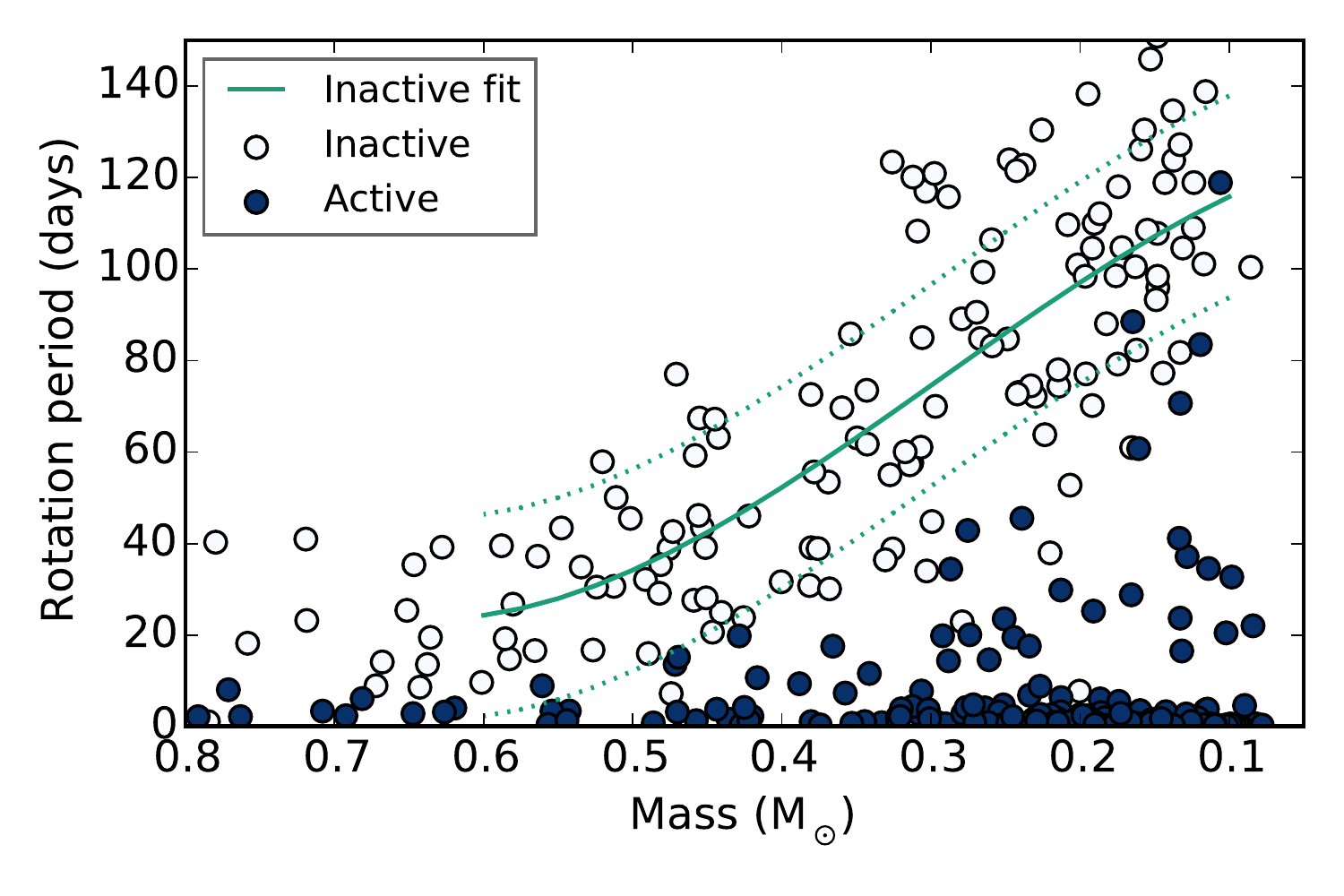}

\caption{Rotation period versus stellar mass for active (filled circles, H$\alpha$ EW $< -1$ \AA) and inactive (white circles, H$\alpha$ EW $> -1$ \AA) stars.  Masses are estimated from a mass--\mk\ relation, which has a scatter of about 10\%. Known or suspected binaries have been removed. The panels differ only in the scaling of the y axis. In the bottom panel, our best-fitting mass--period relation for inactive M dwarfs is also shown (solid line), along with lines indicating the standard deviation in the residuals (dashed lines). \label{Fig:classify}}
\end{figure}

\citet{West2015} noted that for M1V--M4V, all stars rotating faster than $26$ days are magnetically active. For M5V--M8V, a corresponding limit was seen at $86$ days. In Figure \ref{Fig:classify}, we consider the active fraction in light of the mass--period relation. We see a smooth, mass-dependent threshold in whether a star shows H$\alpha$ in emission, with the boundary around $30$ days for $0.3$ \msun\ stars and around $80$ days for $0.15$ \msun. This threshold seems to correspond to the lower boundary of the ``long period'' rotators, which we suggested in \citet{Newton2016} is when an era of rapid angular momentum evolution ceases.

The differentiation of inactive stars at long rotation periods implies that the presence of H$\alpha$ emission is a useful diagnostic for whether a star is a long- or short-period rotator. This may be of use to exoplanet surveys, for which slowly rotating stars are often better targets.
Furthermore, for an inactive star, its mass can be used to provide guidance as to its rotation period. We fit a polynomial between stellar rotation period and mass for inactive stars in our sample, using 3 $\sigma$ clipping to iteratively improve our fit:
\begin{align}
P/\mathrm{days} =& + 127 \\ \nonumber
& - 58\times (M_*/M_\odot) \\ \nonumber
& -587\times (M_*/M_\odot)^2 \\ \nonumber
& + 665 \times (M_*/M_\odot)^3 \nonumber
\end{align}

The relation is valid between $0.1$ and $0.6$ \msun\ and has standard deviation of $22$ days.
The best fit is shown in Figure \ref{Fig:classify}. Note that for early M dwarfs, all but the most rapidly rotating stars are inactive. Because the stars included in this fit are selected only by virtue of being inactive, they are likely to have a range of ages and therefore we do not expect this fit to match up with a particular gyrochrone, or with the Sun.

\subsection{\lhalbol\ saturation level}

\begin{figure}
\includegraphics[width=\linewidth]{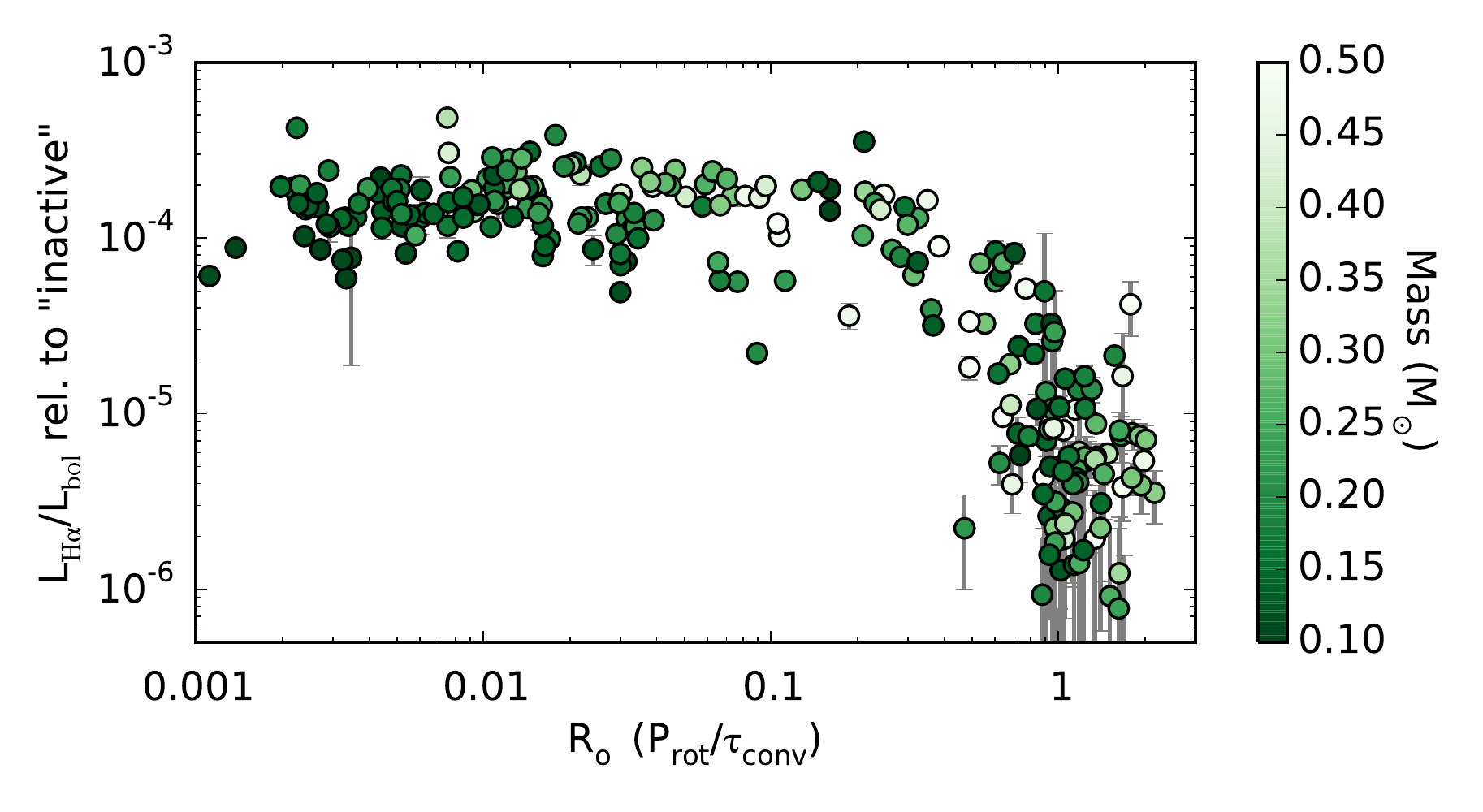}

\caption{\lhalbol\ versus Rossby number (\rossby). We have corrected the H$\alpha$ EWs to be measured relative to the maximum absorption level seen for M dwarfs of the same mass, and used the $\chi$ values from \citet{Douglas2014} to infer \lhalbol\ from EW. For \rossby\, we use the empirical calibration from \citet{Wright2011}. Data points are colored by their estimated stellar mass. We see saturation for rapid rotators (small \rossby), and a decline for slower rotators (large \rossby). \label{Fig:ro_ha}}
\end{figure}

Activity as traced through \lhalbol\ represents the relative amount of the star's luminosity that is output as H$\alpha$ emission and enables a more mass-independent comparison between activity levels in M dwarfs.
The Rossby number (\rossby), which compares the rotation period to convective overturn timescale, is often used to compare activity strengths across mass and rotation period ranges. We use the empirical calibration from \citet{Wright2011} to determine convective overturn timescales. Figure \ref{Fig:ro_ha} shows \lhalbol\ versus \rossby.
We see a saturated relationship between \lhalbol\ and \rossby\ for rapidly rotating stars and a power-law decay in \lhalbol\ with increasing \rossby\ for slowly rotating stars. The break occurs near \rossby$=0.2$.

The mean value in the saturated regime for $M_*<0.25$ \msun\ is $(1.536\pm0.004)\times 10^{-4}$. This is lower than the saturation value for $M_*>0.25$ \msun, which is ($1.852\pm0.007)\times 10^{-4}$. The errors here represent the standard error on the mean. \edit1{Note that if we considered \lhalbol\ without adjusting for the mass-dependent H$\alpha$ absorption as done in other works, our mean values would be about 0.1 dex lower for the early M dwarfs and 0.04 dex lower for the late Ms. We also note that the median value is lower than the mean for the low-mass M dwarfs ($1.428\times 10^{-4}$).} 

\citet{West2004} and \citet{Kruse2010} show a similar decrease in \lhalbol\ for spectral types M6 and later, and similar levels of mean \lhalbol. The range of values we find is consistent with that seen in other studies of field stars, for example \citet[][Fig. 8]{Gizis2002} and \citet[][Fig. 9]{Reiners2012a}. It is higher than the saturation threshold recently reported for the Hyades \citep[\lhalbol=$1.26\pm0.04 \times 10^{-4}$;][]{Douglas2014}, but differences in analysis technique have not been addressed and there is significant intrinsic scatter to the saturation level.

\subsection{Activity versus Rossby number}

\begin{figure}
\includegraphics[width=\linewidth]{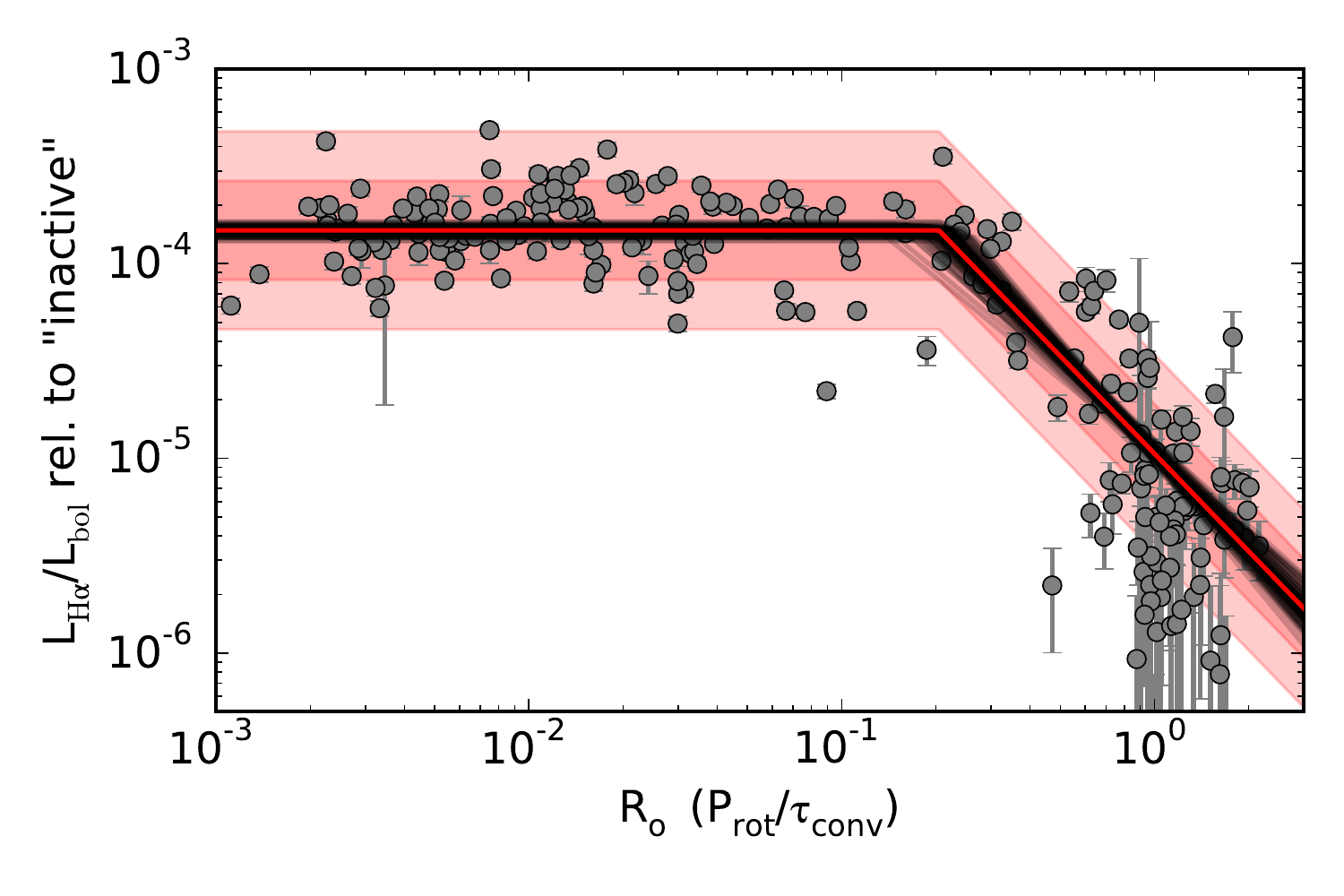}

\caption{\lhalbol\ versus Rossby number (\rossby), as in Figure \ref{Fig:ro_ha}. We fit the canonical rotation-activity relation to our data, with \lhalbol\ maintaining a constant value for small \rossby\ and a power-law decay in \lhalbol\ at larger \rossby. Our best fit is shown as the solid red line, with 100 random draws from the posterior distribution shown in black. We also fit for intrinsic scatter $\sigma_\mathrm{LHa}$ within the rotation-activity relation; the $1\sigma_\mathrm{LHa}$ and $2\sigma_\mathrm{LHa}$ contours are indicated by the shaded red region. \label{Fig:fits}}
\end{figure}

We fit the canonical activity-rotation relation to our data:
\begin{equation}
L_{H\alpha}/L_{bol} = 
\begin{cases}
\left( L_{H\alpha}/L_{bol} \right)_{sat}, & R_o \leq R_{o,sat} \\
C R_o^{\beta}, & R_o>R_{o,sat}
\end{cases}
\end{equation}
We use the open-source Markov chain Monte Carlo sampler package \texttt{emcee} \cite{Foreman-Mackey2013} for this analysis. This allows us to include intrinsic scatter in the relation, which we model as a constant amount of scatter $\sigma_\mathrm{LHa}$ on $\log{L_{H\alpha}/L_{bol}}$ \citep[see e.g.][]{Hogg2010}. We use uninformative uniform priors on $\beta$, $\log{\left( L_{H\alpha}/L_{bol} \right)_{sat}}$, $\log{R_{o,sat}}$, and $\log{\sigma}$. The acceptance rate is $0.6$, and the autocorrelation timescale is 20-50 steps. Therefore, following the recommendations of \citet{Foreman-Mackey2013}, we run $200$ walkers for a total of $700$ steps, discarding the first $200$ as burn-in. We take the median as the best-fitting values and report error bars corresponding to the 16$^{\rm th}$ and 84$^{\rm th}$ percentiles on the marginalized distributions:
\begin{align*}
&R_{o,sat} = 0.21 \pm 0.02\\
&L_{\mathrm{H}\alpha}/L_{\mathrm{bol}_{sat}} = (1.49 \pm 0.08) \times 10^{-4} \\
&\beta = -1.7 \pm 0.1 \\
&\sigma_\mathrm{LHa} = 0.26 \pm 0.01
\end{align*}
Figure \ref{Fig:fits} shows the best fit, including intrinsic scatter, over-plotted on our data. Also shown are 100 random draws from the posterior probability distribution. Note that $\sigma_\mathrm{LHa}$ incorporates both intrinsic variation at a given stellar mass and the difference in saturation level between early and late M dwarfs.

\begin{figure}
\includegraphics[width=\linewidth]{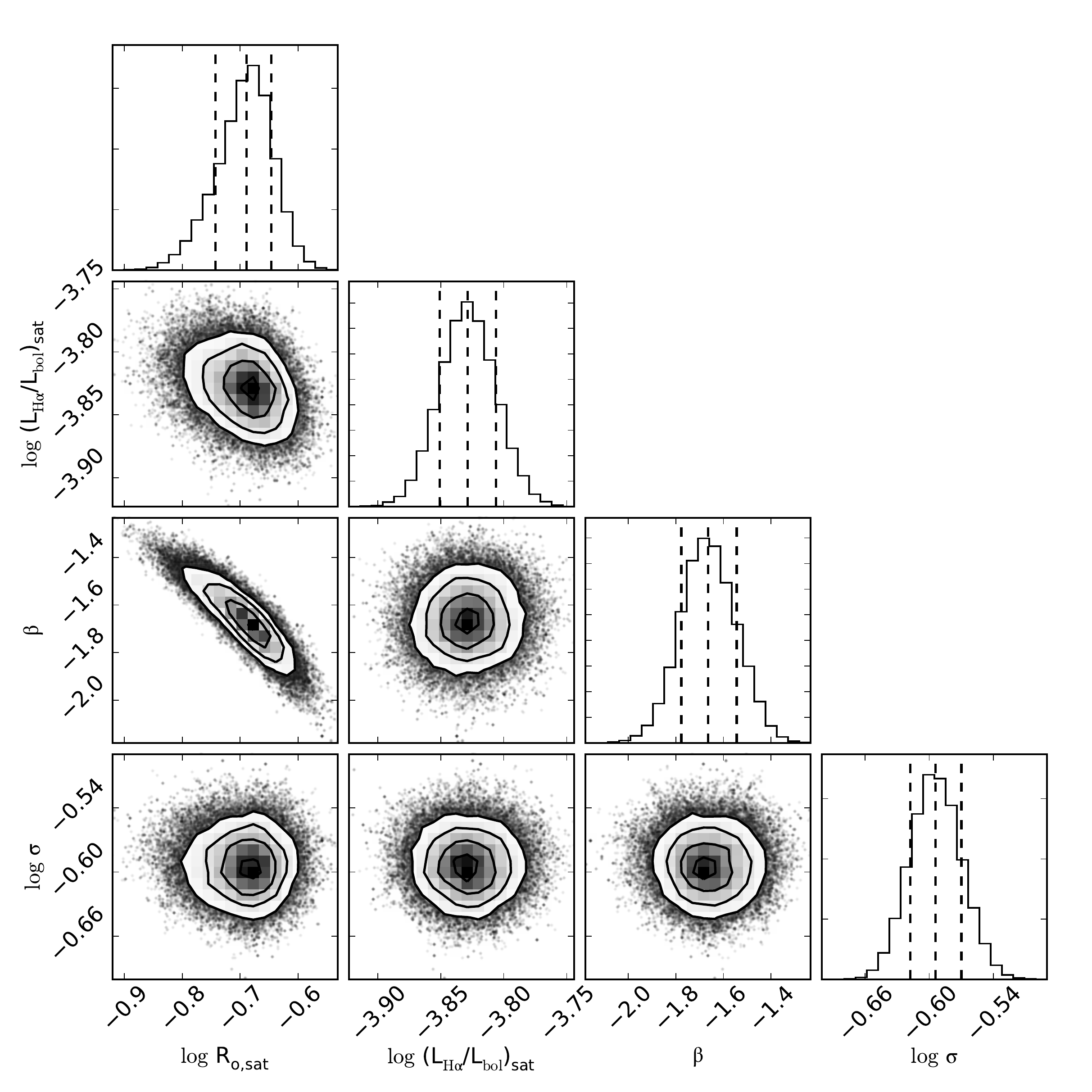}

\caption{The posterior distributions over each parameter in our rotation-activity relation. Contours are shown at $0.5$, $1$, $1.5$, and $2\sigma$. The marginalized distributions for each parameter (histograms along the diagonal) are shown, with the 16$^{\rm th}$ and 84$^{\rm th}$ percentiles and the median indicated as dashed lines. \label{Fig:corners}}
\end{figure}

Figure \ref{Fig:corners} shows the posterior probability distributions over each parameter \citep[using the \texttt{corner} package;][]{Foreman-Mackey2016}. As in \citet{Douglas2014}, there is an anti-correlation between $\beta$ and $R_{o,sat}$. Our results are marginally inconsistent with \citet{Douglas2014}, who obtained $(L_{H\alpha}/L_{bol})_{sat}=(1.26\pm0.04) \times 10^{-4}$, $R_{o,sat}=0.11^{+0.02}_{-0.03}$, and $\beta=-0.73^{+0.16}_{-0.12}$ for M dwarfs in the Hyades and Praesepe. The strong degeneracy between $\beta$ and $R_{o,sat}$ and the unaccounted for mass-dependence of the saturation level are potential contributors.

\subsection{Stars with unusual activity levels}\label{Sec:oddballs}

\edit1{One star appears as an outlier in the mass--period--activity plane: 2MASS J23242652+7357437 (LP 48-485) is a rapid rotator and yet is inactive. The H$\alpha$  EW of $-0.52\pm0.04$ is from this work, and the rotation period of $7.738$ days is from \citet{Newton2016}.}

\edit1{We obtained two high resolution spectra of 2MASS J23242652+7357437 using TRES, which revealed no change in RV and no obvious evidence of a second set of lines. A small amount of H$\alpha$ emission is seen in the TRES spectrum. We re-examined the light curve from which \citet{Newton2016} derived the rotation period and found no alternative rotation periods.}

Though unusual, there are three stars analyzed by \citet{West2009} that indicate the existence of rapidly rotating, inactive M dwarfs. \citet{West2009} suggested complexity in the rotation--activity relation as the cause. However, we have demonstrated a clear connection between rotation and H$\alpha$ activity for M dwarfs of all masses, which makes this ``oddball'' star even more puzzling. 

\edit1{Several additional targets initially appeared in this list of oddball stars, but further investigation with extant photometry and new high-resolution spectroscopic measurements showed either that the stars were binaries or that there were issues with the assumed rotation periods (see \S\ref{Sec:data}). We note that we did not consider stars flagged as binaries in our search for ``oddballs'' due to the uncertainty over which companion is the source of the rotation and/or H$\alpha$ signal and the potential for binarity influencing evolution. }

\subsection{Activity versus photometric amplitude}

Photometric rotational modulation results from starspots rotating in and out of view on the stellar surface. Since starspots are the product of the magnetic suppression of flux, we might expect a correlation between the prevalence of starspots and spectral indicators of magnetic activity. The photometric rotation amplitude is indicative of the fraction of the stellar surface that is covered in spots, though it is primarily sensitive to asymmetries in the longitudinal distribution of starspots. 

\begin{figure}
\includegraphics[width=0.99\linewidth]{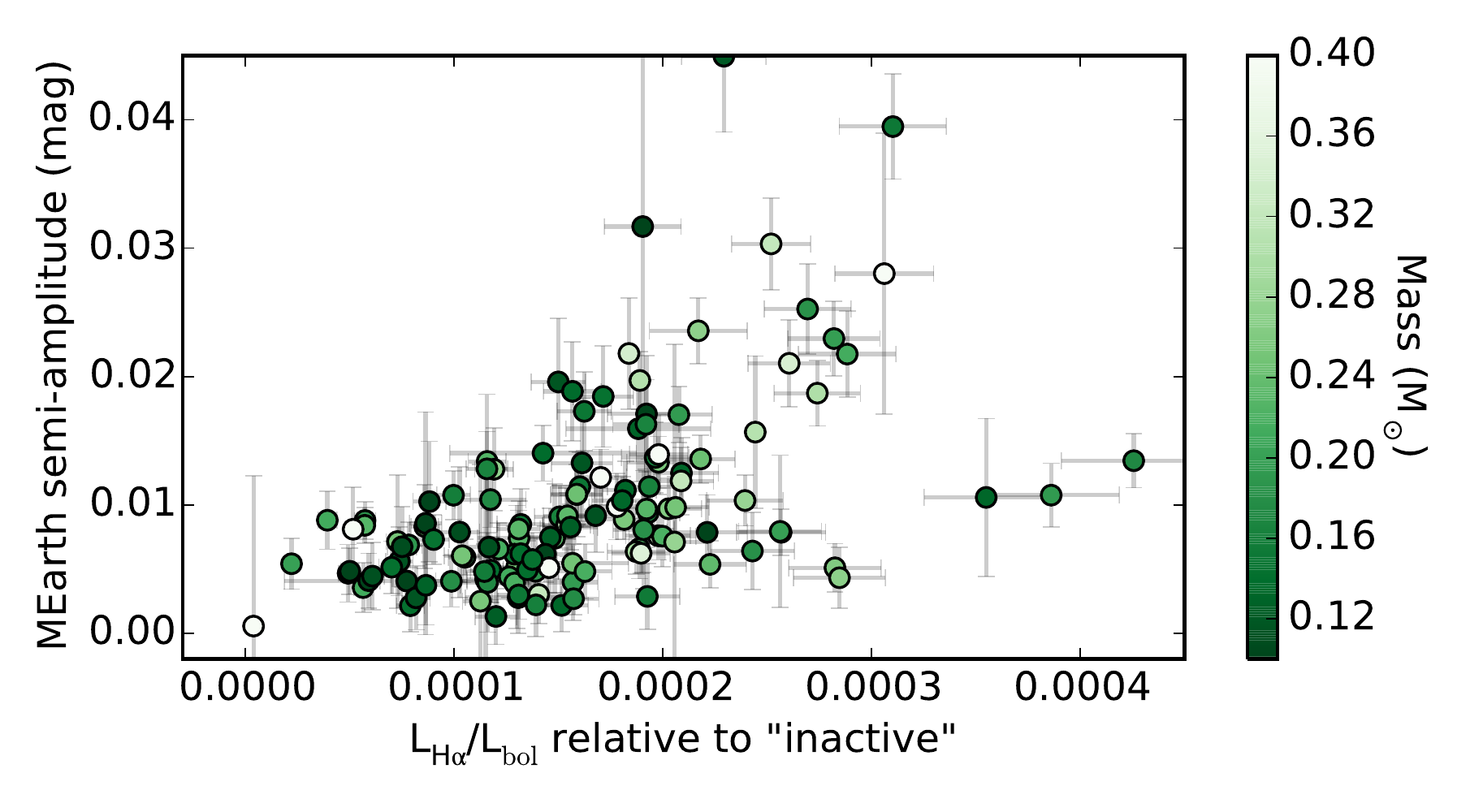}

\includegraphics[width=\linewidth]{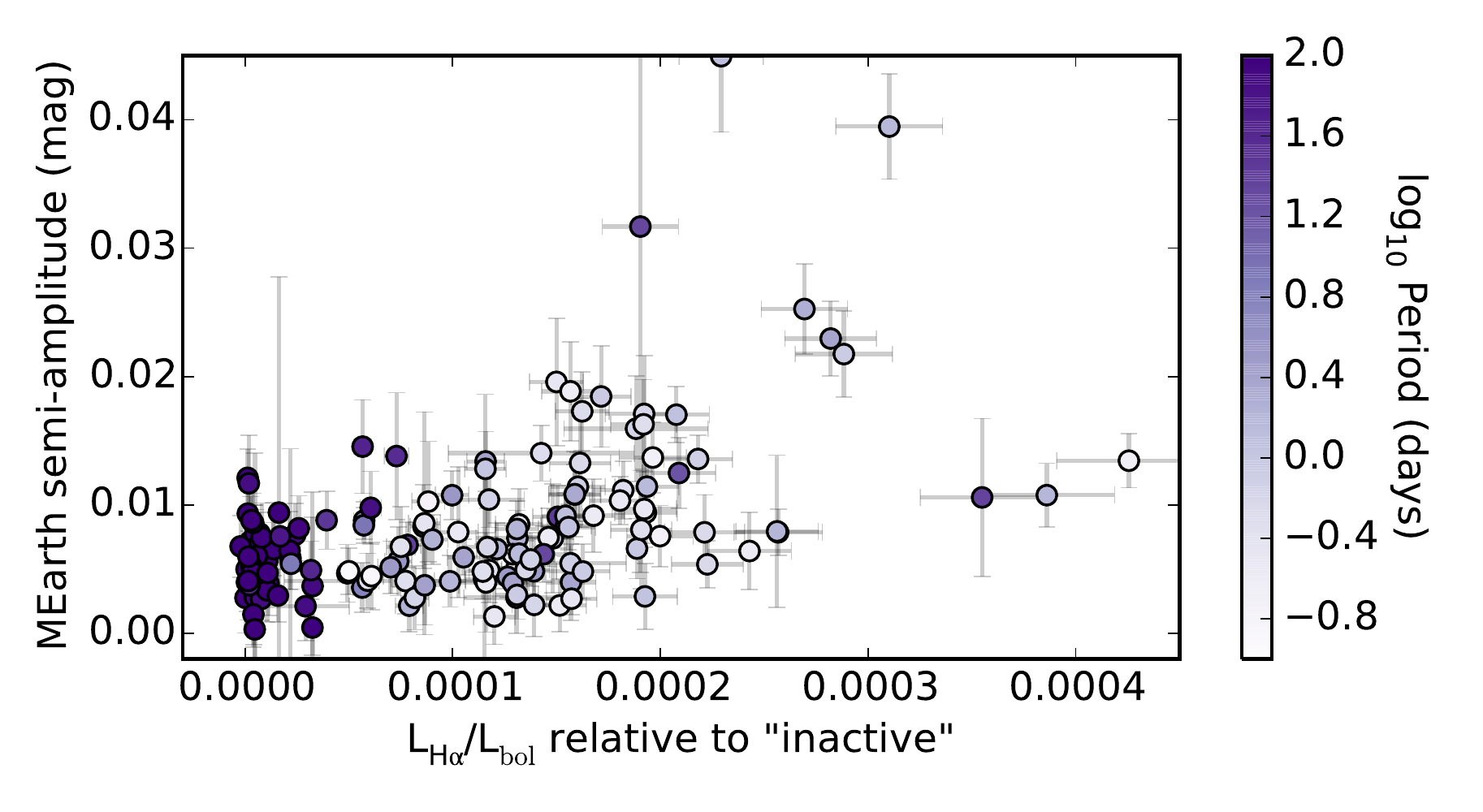}

\caption{Amplitude of photometric variability versus \lhalbol\ for M dwarfs with detected rotation periods from \citet{Newton2016}. The mean amplitude in different bins in \lhalbol\ is shown, along with the error on the mean. In the top panel, only stars with periods faster than $30$ days are shown, and the color of the data point indicates stellar mass. In the bottom panel, only stars with masses less than $0.25$ \msun\ are shown, and the color of the data point indicates rotation period. In both panels, a highly significant correlation between amplitude and magnetic activity is seen. \label{Fig:amp}}
\end{figure}

For \lhalbol\ greater than about $1\times10^{-4}$, we see an increase in the dispersion of photometric amplitudes with increasing \lhalbol, such that the stars displaying the highest amplitude of variability are also the strongest H$\alpha$ emitters (Figure \ref{Fig:amp}). This is evidenced by a strong positive correlation between the two parameters. The Spearman rank correlation coefficient $\rho$ is $0.39\pm0.03$, with a $p$-value of $<4\times10^{-8}$. The most active and highly variable stars contribute to the strength of the correlation, but a correlation persists if we exclude stars with \lhalbol$>2.5\times10^{-4}$ ($\rho=0.36\pm0.04$, $p<0.0002$).

One potential concern is that we found that active stars are slightly redder than inactive stars, resulting in $\chi$ values for the active sample that are $5\%$ lower. However, this has the opposite effect of the observed correlation:
if we were to assign $\chi$ values on the basis of spectral type rather than color, the active stars of that spectral type would be assigned a larger $\chi$ than otherwise, and would therefore have larger \lhalbol. We nevertheless verified that our results are unchanged if we use H$\alpha$ EW in place of \lhalbol\ ($\rho=-0.4\pm0.3$, $p<3\times10^{-5}$). 

Since photometric amplitudes depend on the bandpass, we only use stars with rotation period measurements from our analysis of MEarth photometry \citep{Newton2016}. One concern is that the method we used to determine amplitude tends to suppress amplitude for stars with strong spot evolution or with non-sinusoidal variability; our method for period detection is also most sensitive to stars with stable, sinusoidal spot patterns. Measuring the peak-to-peak amplitude offers an alternative, and was used, e.g.\ by \citet{McQuillan2014} in their study of rotation in the \kepler\ sample. However, this is a more challenging measurement to make robustly in ground-based data, particularly if there are gaps in phase coverage. We therefore proceeded with the amplitude measurements from \citet{Newton2016}.

Another complicating factor is the relationship between rotation period and photometric amplitude, since the former is also correlated with H$\alpha$ activity. For stars more massive than $0.25$ \msun, a negative correlation is seen between variability amplitude and rotation period for periods $>30$ days \citep{Hartman2011, McQuillan2014, Newton2016}. However, for the mid-to-late M dwarfs that dominate our sample, no correlation is seen \citep{Newton2016}. To address this concern, we performed our analysis on different subsets of data, which are shown in Figure \ref{Fig:amp}. In the first, we have restricted the period range to be $<30$ days; in the second we have restricted the mass range to be $<0.25$ \msun. The results from each restricted sample are consistent, with $\rho=0.41\pm0.04$ for the period-restricted sample and $\rho=0.34\pm0.03$ for the mass-restricted sample, both with $p<10^{-4}$. Restricting on \rossby\ produces consistent results.

There are three stars in our sample with \lhalbol $>3.5\times10^{-4}$. The amplitudes of their rotational modulations are smaller than what is seen in slightly less active stars (we verified by eye that the amplitudes of these modulations are not artificially low due to spot evolution). If real, this trend may reflect an increase in the filling factor of spots, with spotted surface now dominating the unspotted surface.

\section{Summary and discussion}

We have obtained new optical spectra for \numnew\ nearby M dwarfs and measured H$\alpha$ EWs and estimated \lhalbol. Including measurements compiled from the literature, our sample includes \numall\ measurements of or upper limits on H$\alpha$ emission. Of these, \numrot\ have photometric rotation periods. These period measurements are primarily from our analysis of data from the MEarth-North observatory \citep{Newton2016}. High-quality M dwarf light curves are available from space-based data and offer unique opportunities for studying M dwarf stellar physics \citep[e.g.][]{Hawley2014, Stelzer2016}, but the disadvantage is that the sample is dominated by early M dwarfs and stars that are typically more distant and therefore not as well characterized as Solar Neighborhood stars. Our sample comprises both early and late M dwarfs within the Solar Neighborhood, most with trigonometric distances and spectroscopic follow-up, and includes around $300$ stars whose masses indicate that there are below the fully convective boundary.

\subsection{The rotation-activity relation}

We have shown that with very high confidence, an M dwarf without detectable H$\alpha$ emission is slowly rotating. For inactive M dwarfs, we have presented a relationship between stellar mass and rotation period. These findings may be useful to those building target lists for exoplanet surveys, providing a simple  and accessible diagnostic of the stellar rotation period. We also suggest that, in the eventuality that gyrochronology is calibrated for M dwarfs, the lack of H$\alpha$ emission can be used to determine whether it is appropriate to apply the gyrochronology relationship.

We fit \lhalbol\ as a function of \rossby\ using the canonical rotation-activity relation, which consists of a saturated regime and one described by a power-law decay. Our photometric rotation periods allow us to investigate the latter part of the relation, where the rotation period becomes comparable to, or longer than, the convective overturn timescale (\rossby\ $\simeq$1) for a range of stellar masses. In M dwarfs, this regime is inaccessible when using \vsini\ as a tracer of rotation. 

For rapidly rotating stars, H$\alpha$ emission maintains a saturated value, as seen in many previous works \citep[][]{Delfosse1998, Mohanty2003, Reiners2012a, Douglas2014}. The saturation value for early M dwarfs is lower for low-mass M dwarfs than it is for high-mass M dwarfs. For \rossby\ $>0.2$, the decline in \lhalbol\ has a power law index of $-1.7\pm0.1$.
Around \rossby $=1$, H$\alpha$ has diminished to the point where it is not detectable in emission in our low-resolution spectra (note, however, that by correcting our H$\alpha$ EWs such that they are measured relative to a maximum absorption level, we obtain a measure of relative \lhalbol\ for  \rossby\ as large as 2).

\begin{figure}
\includegraphics[width=\linewidth]{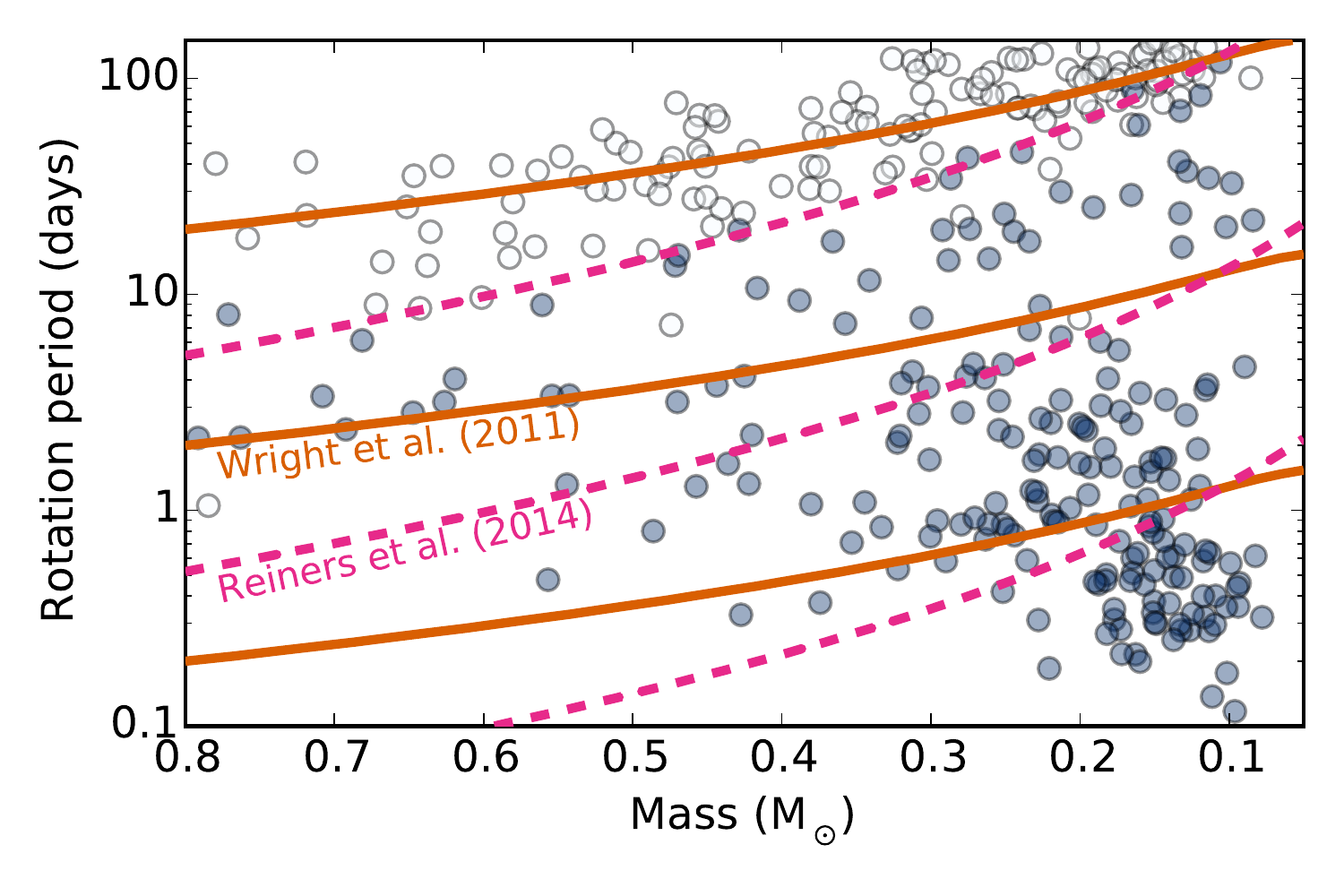}

\caption{Stellar rotation versus mass, showing active (filled) and inactive (open) stars. Over-plotted are contours of constant \rossby\ \citep[solid lines; using the empirical calibration from][]{Wright2011}, and the generalized \rossby\ scaling from \citet[][dashed lines]{Reiners2014}. For the former, the contours are $0.01$, $0.1$ and $1$ from bottom to top. For the latter, they are at $1000/k$, $10/k$, and $0.1/k$ from bottom to top. \label{Fig:rossby}}
\end{figure}

\begin{figure}
\includegraphics[width=\linewidth]{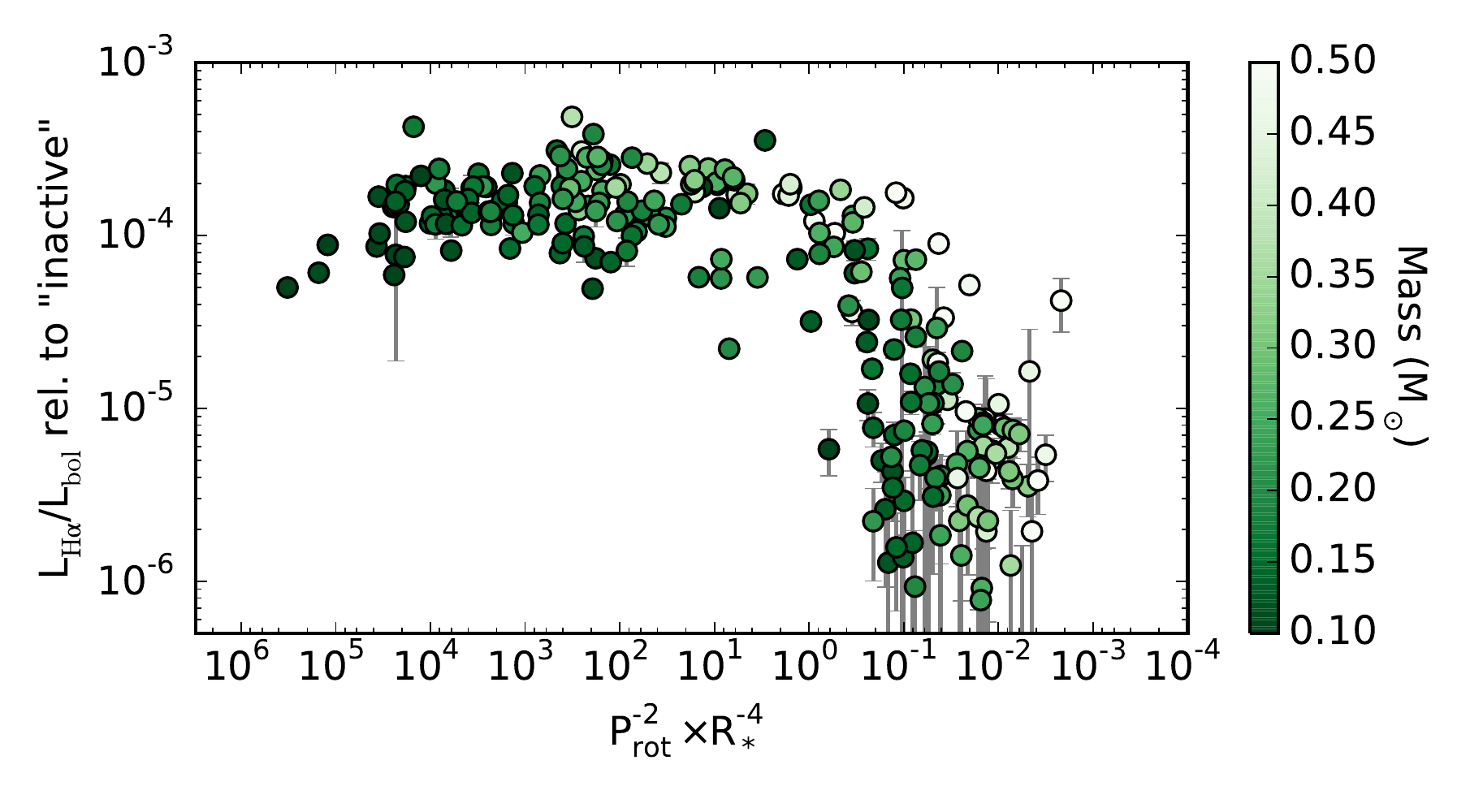}

\caption{The activity-rotation relation, similar to Figure \ref{Fig:ro_ha}, but using the ``generalized \rossby'' scaling from \citet{Reiners2014} instead of the \rossby\ from \citet{Wright2011}. Using the R14 scaling, the slope in the unsaturated regime is a mass-dependent. \label{Fig:reiners}}
\end{figure}

\citet[][hereafter R14]{Reiners2014} suggest that \rossby\ is not the best scaling, and explored a generalized relationship between \lhalbol\ and rotation period and stellar radius. They find $L_\mathrm{X}/L_\mathrm{bol} = k P^{-2} R^{-4}$. In Figure \ref{Fig:rossby}, we show how the R14 scaling and \rossby\ numbers differ in the mass--period plane. The former depends on stellar radius, so we use the mass--radius relation from \citet{Boyajian2012} to estimate stellar radius. Figure \ref{Fig:reiners} shows \lhalbol\ versus the R14 scaling in place of \rossby. The R14 scaling matches the shape of the long-period sample very well, with $L_\mathrm{X}/L_\mathrm{bol} = 0.1 P^{-2} R^{-4}$ (the $0.1/k$ contour) aligning with the active/inactive boundary. In contrast to \lhalbol\ vs. \rossby, we find that with the R14 scaling, the slope in the unsaturated regime is dependent on stellar mass.

\subsection{The amplitude-activity relation}

Considering stars with detected photometric rotation periods, we have found that more highly variable stars are also more active. This is seen through a highly significant correlation between the strength of H$\alpha$ emission and the amplitude of photometric variability. Both starspots and H$\alpha$ emission are thought to be products of magnetism. We expect that this correlation is the result of differences in the underlying magnetic field strength: stars with stronger magnetic fields have stronger H$\alpha$ emission as well as larger or more abundant spots. 

On the other hand, \citet{Jackson2012} looked at stars in the open cluster NGC 2516, which had been surveyed photometrically as part of the Monitor program \citep[]{Irwin2007}. They found no difference in the chromospheric activity between the stars with and without rotation period measurements, and argued that there were not differences in the spot filling factor between these two groups. 

The amplitude-activity trend may be related to the difference in $\chi$ values for active and inactive M dwarfs: active M dwarfs are slightly redder than inactive M dwarfs, perhaps resulting from a greater abundance of cool spots on active stars. \edit1{A similar effect was seen in the broadband colors of Pleiades K and M dwarfs \citep{Stauffer2003, Kamai2014, Covey2016}. Rotationally variable stars in the Pleiades were found to be redder in $V-K$ (suggested to be the result of starspots), and bluer in $B-V$ (suggested to be the result of plages). \citet{Stauffer2003} note that this would require spot filling factors of $\gtrsim50\%$.}

Inferred spot filling factors for M dwarfs range from on the order of a few percent \citep{Barnes2015, Andersen2015} to 40\% \citep[e.g.][]{Jackson2013}, though measurements are likely complicated by the unknown spot temperature and geometry. For small randomly distributed spots with filling factors $<20\%$, simulations from \citet[][see Fig.~5]{Andersen2015} indicate that an increase in the filling factor by a factor of two corresponds to a $50\%$ increase in photometric variability in $V$. We see a factor of two increase in photometric variability (in the MEarth red-optical bandpass, where the spot contrast is diminished) between the highly active and the inactive stars, which would require a several-times increase in the filling factor. 
Alternatively, M dwarfs may be dominated by one or more larger spots. 

The three most active stars in our sample have small variability amplitudes, contrary to the trend we see in the less active stars. If this result holds with the inclusion of additional data, it could indicate that spots are covering more than $50\%$ of the stellar surface in these highly active stars.

\subsection{Implications for the magnetic dynamo}

We see a clear mass-dependent rotation period threshold for H$\alpha$ emission. In \citet{Irwin2011} and \citet{Newton2016}, we found a dearth of mid-to-late M dwarfs with intermediate rotation periods, which we suggested represents a period range over which stars quickly lose angular momentum. The active/inactive boundary coincides with the rotation period at which this rapid evolution appears to cease, suggesting a connection between the two phenomena. 

The stars in our sample span the fully convective boundary, covering the full range of expected rotation periods between masses of $0.5$ and $0.1$ solar masses. A gradual change in magnetic dynamo is expected over this regime due to the diminishing radiative zone, with the disappearance of the tachocline occurring around $M_*=0.35$\msun.  Other than a difference in the saturation level of \lhalbol\ between early and late M dwarfs, we find a single relationship between \lhalbol\ and \rossby\ for all stars in our sample.  It could be that the magnetic heating of the chromosphere (as traced through H$\alpha$) is independent of the underlying magnetic dynamo, or the change in the magnetic dynamo across this mass range may be not sufficiently dramatic as to manifest in an intrinsically variable tracer like H$\alpha$.

Alternatively, the magnetic dynamo may \emph{not} change across this mass range. \citet{Wright2016} suggest a common magnetic dynamo in solar-type and fully convective stars. They found that the relationship between x-ray luminosity and \rossby\ relation in fully convective stars resembles that of solar-type stars \citep[see also][]{Kiraga2007, Jeffries2011}. We note that the sample of fully convective stars in the unsaturated regime with x-ray measurements is much smaller than the sample with available H$\alpha$ measurements.

However, is important to recall that mass-normalizations are part of both our \lhalbol\ measure (which is measured relative to the maximum absorption observed at a given stellar mass and includes a color-dependent conversion from EW to luminosity) and \rossby\ (which is empirically derived to minimize scatter in the x-ray--\rossby\ plane). The lack of observed mass dependence in \lhalbol--\rossby\ relation could signify that the \rossby\ number sufficiently accounts for the gradual changes in the magnetic dynamo expected as the fully convective boundary is crossed. For example, the empirical convective overturn timescale ($\tau_\mathrm{conv}$) from \citet{Wright2011} has a sharp increase in slope around $0.35$ \msun. We also see mass dependence when using the scaling relation suggested by \citet{Reiners2014} in place of \rossby. 

\edit1{It is nevertheless readily apparent from our data that rotation plays a critical role in determining the H$\alpha$ emission strength -- and by extrapolation the magnetic dynamo -- of both partially and fully convective M dwarfs.}

\appendix

\section{Absolute radial velocities in low-resolution optical spectra}\label{Sec:rv}

\subsection{Method}

To measure radial velocities (RVs) in our FAST spectra, we forward model the velocity shift and the difference in shape between the science spectrum and an RV standard. For accurate RVs, this analysis requires a close match between the science and standard spectra. This is due to the complex molecular absorption features in M dwarf spectra, which change significantly across the M spectral class, with noticeable differences even between adjacent spectral types. We use a $5^{th}$-order Legendre polynomial to account for the continuum mismatch \citep[e.g][]{Anglada-Escude2012b}. We use linear least squares to determine the coefficients of the Legendre polynomial at a grid of velocity shifts, producing a $\chi^2$ value at each test velocity. We then fit a parabola to velocity shifts with $\chi^2$ values that are within 1\% of the lowest $\chi^2$, and adopt the vertex of the parabola as the best fitting radial velocity.

Despite fitting for the difference in shape between the science spectrum and RV standard, we found that a mismatch between the spectral types could result in systematic differences of a few to $10$ km/s. Therefore, we measure the RV against a standard of each M spectral type. We then select the standard and RV that resulted in the lowest $\chi^2$. The spectral types selected using this technique generally agrees to within one spectral type of those assigned by eye in \citet{West2015}.

For our RV standards, we adopt the zero-velocity SDSS spectra from \citet{Bochanski2007a}\footnote{Available on github: \url{https://github.com/jbochanski/SDSS-templates}}, which we convert to air wavelengths. We masked the red edge of spectra ($\lambda > 7300$ \AA), $20$ \AA\ regions surrounding the Na D and H$\alpha$ lines, the oxygen B band ($6860<\lambda<6970$ \AA), and the water band ($7160<\lambda<7340$ \AA). We use the spectra where highest quality active and inactive stars have been averaged.

\subsection{Accuracy and precision}

We assess the accuracy of our absolute radial velocities by comparing our measured values to those from \citet{Gizis2002}, \citet{Delfosse1998}, \citet{Nidever2002}, and \citet{Newton2014}. Note that the absolute rest frame we used in \citet{Newton2014} is based on comparison to \citet{Nidever2002}, so these comparisons are not independent. Table \ref{Tab:rv_comp} summarizes our results. We find a systematic offset between our measured RVs and those from each of these studies, with a mean offset (this work - literature) of $+7$ to $+11$ km/s. 
There is not a significant difference between our values and those from \citet{West2015}, which suggests that the offset is present in that work. This is consistent with the $7.3$ km/s offset has previously been identified in the SDSS-SEGUE sample \citep{Yanny2009}, so we correct our RVs by applying this offset.

However, we caution that there may be additional, spectral-type dependent systematic errors. For example, when using the SDSS active-only templates or inactive-only templates, we found systematic offsets on the order of $4$ km/s for M4V-M6V stars. \edit1{This is important for galactic kinematics and we caution against using the RVs we present for such a purpose. For our present goal of measuring H$\alpha$ emission strength,  the differences in RVs discussed above are negligible: the resolution of our spectra is about $100$ km/s.}

The standard deviation of the difference between our RVs and those from \citet{West2015}, which are based on the different reductions of the same data, indicates the precision with which the radial velocity can be determined in the absence of noise. We find this to be $3$ km/s. However, the typical random error will be larger. Comparing to \citet{Newton2014}, the standard deviation in the RV difference is $9$ km/s. In \citet{Newton2014}, the typical error was about $4.5$ km/s. \edit1{We assume that the errors in the FAST RVs and the errors from \citet{Newton2014} can be added in quadrature to produce the $9$ km/s standard deviation in the RV difference. This means that the FAST RV errors are about $8$ km/s, which we adopt as the random error on our measurements.}

\begin{deluxetable*}{l l r r r}
\tablecaption{Comparison of RV measurements \label{Tab:rv_comp}}
\tablehead{
\colhead{Reference} &
\colhead{dM Sp. Type} &
\colhead{$N_\mathrm{stars}$} &
\colhead{$\mean{\Delta V}$} &
\colhead{$\sigma_{\Delta V}$} }
\startdata
\citet{Gizis2002} & 1-7 & 44 & $+9.7\pm1.1$ & 7.4 \\ 
$\cdots$ & 1 & 7 & $+3.6\pm2.1$ & 5.5 \\ 
$\cdots$ & 3 & 6 & $+11.5\pm2.0$ & 5.0 \\ 
$\cdots$ & 4 & 11 & $+12.3\pm2.2$ & 7.5 \\ 
$\cdots$ & 5 & 13 & $+7.4\pm1.7$ & 6.2 \\ 
\citet{Delfosse1998} & 1-7 & 20 & $+11.5\pm1.5$ & 6.8 \\ 
\citet{Nidever2002} & 1-5 & 15 & $+8.3\pm1.2$ & 4.6 \\ 
\citet{Newton2014} & 1-8 & 207 & $+6.8\pm0.6$ & 8.9 \\ 
$\cdots$ & 3 & 10 & $+5.0\pm1.9$ & 6.0 \\ 
$\cdots$ & 4 & 40 & $+11.0\pm1.5$ & 9.7 \\ 
$\cdots$& 5 & 95 & $+6.8\pm0.8$ & 7.6 \\ 
$\cdots$& 6 & 46 & $+5.0\pm1.5$ & 10.4 \\ 
$\cdots$ & 7 & 11 & $+3.2\pm2.3$ & 7.5 \\ 
\citet{West2015} & 1-8 & 207 & $-0.5\pm0.3$ & 3.7 \\ 
$\cdots$ & 3 & 7 & $-3.7\pm1.7$ & 4.5 \\ 
$\cdots$ & 4 & 41 & $+0.1\pm0.5$ & 3.5 \\ 
$\cdots$ & 5 & 93 & $-1.1\pm0.3$ & 2.9 \\ 
$\cdots$ & 6 & 41 & $+1.0\pm0.5$ & 3.2 \\ 
$\cdots$ & 7 & 18 & $-1.6\pm1.4$ & 5.8 
\enddata
\tablenotetext{a}{The measurements from \citet{West2015} and this work are based on different reductions of the same data and different RV codes.}
\tablecomments{All RVs are in km/s. $\Delta V$ is defined as the mean of $V_\mathrm{ref} - V_\mathrm{FAST}$. Comparisons to five different surveys are reported. We also report differences with the comparison limited to a single spectral type, when there are at least five stars of that spectral type available (spectral types are determined using the $\chi^2$ spectral types we derive in this work). }
\end{deluxetable*}

\acknowledgments We thank E. Mamajek, E. Berger, D. Latham, and C. Conroy for useful discussions. ERN is supported by an NSF Astronomy and Astrophysics Postdoctoral Fellowship under award AST-1602597. The MEarth project acknowledges funding from the National Science Foundation under grants AST-0807690, AST- 1109468, and AST-1004488 (Alan T. Waterman Award) and the David and Lucile Packard Foundation Fellowship for Science and Engineering. This publication was made possible through the support of a grant from the John Templeton Foundation. The opinions expressed here are those of the authors and do not necessarily reflect the views of the John Templeton Foundation. This research has made use of data products from the Two Micron All Sky Survey, which is a joint project of the University of Massachusetts and the Infrared Processing and Analysis Center / California Institute of Technology, funded by NASA and the NSF; NASA Astrophysics Data System (ADS); and the SIMBAD database and VizieR catalog access tool, at CDS, Strasbourg, France.

\clearpage

\end{document}